 \documentclass[10pt,twocolumn,twoside]{IEEEtran} 

\usepackage{graphicx, amsmath, amsfonts}
\usepackage{framed}
\usepackage{algorithm}
\usepackage{algorithmic}
\usepackage{lipsum,amsmath,multicol}
\usepackage[update,prepend]{epstopdf}
\usepackage{subfigure}
\usepackage{cite}
\usepackage{latexsym}
\usepackage{amssymb}
\usepackage{afterpage}

\newtheorem{theorem}{Theorem}
\newtheorem{corollary}{Corollary}
\newtheorem{proposition}{Proposition}
\newtheorem{lemma}{Lemma}
\newtheorem{definition}{Definition}

\IEEEoverridecommandlockouts

\begin{document}

\title{Random Subcarrier Allocation \\ in OFDM-Based Cognitive Radio Networks}

\author{Sabit~Ekin,~\IEEEmembership{Student~Member,~IEEE,}~Mohamed~M.~Abdallah,~\IEEEmembership{Member,~IEEE,} Khalid~A.~Qaraqe,~\IEEEmembership{Senior~Member,~IEEE,}~and~Erchin~Serpedin,~\IEEEmembership{Senior~Member,~IEEE}

\thanks{Manuscript received Dec. 10, 2011; revised Mar. 23, 2012; accepted Apr. 20, 2012. Date of publication XXXX XX, 20XX; date of current version XXXX XX, 20XX. The associate editor coordinating the review of this manuscript and approving it for publication was Dr. Sofiene Affes. This work was made possible by the support offered by NPRP grants No. 09-341-2128, 08-152-2-043 and 4-1293-2-513 from the
Qatar National Research Fund (a member of Qatar Foundation). This paper was presented in part at the Wireless Advanced Conference, London, UK, June 2012.}
\thanks{Copyright (c) 2012 IEEE. Personal use of this material is permitted. However, permission to use this material for any other purposes
must be obtained from the IEEE by sending a request to pubs-permissions@ieee.org.}
\thanks{S. Ekin and E. Serpedin are with the Department of Electrical and Computer Engineering, Texas A\&M
University, College Station, TX 77843-3128 (e-mail: sabitekin@gmail.com; serpedin@ece.tamu.edu).}
\thanks{M. M. Abdallah and K. A. Qaraqe are with the  Department of Electrical and Computer Engineering, Texas A\&M
University at Qatar, Education City, Doha, Qatar (e-mail: \{mohamed.abdallah; khalid.qaraqe\}@qatar.tamu.edu).}
}

\markboth{IEEE Transactions on Signal Processing,~Vol.~XX, No.~XX, XXX~2012}%
{Ekin \MakeLowercase{\textit{et al.}}: Random Subcarrier Allocation in OFDM-Based Spectrum Sharing Wireless Networks }

\maketitle

\begin{abstract}
This paper  investigates the performance of an orthogonal frequency-division multiplexing (OFDM)-based cognitive radio (CR) spectrum sharing communication system that
assumes  {\it random allocation}  and absence  of the primary user's (PU) channel occupation information, i.e., {\it no spectrum sensing} is employed to acquire information about the availability of unused subcarriers.
In case of a single secondary user (SU) in the secondary network, due to the lack of information of PUs' activities,  the SU randomly allocates the subcarriers of the primary network and {\it collide} with the PUs' subcarriers with a certain probability. To maintain the quality of service (QoS) requirement of PUs, the interference that SU causes onto PUs is controlled by adjusting SU's transmit power below a predefined threshold, referred to as interference temperature. In this work, the average capacity of SU with subcarrier collisions is employed as performance measure to investigate the proposed random allocation scheme for both general and Rayleigh channel fading models. Bounds and scaling laws of average capacity with respect to the number of SU's, PUs' and available subcarriers  are derived. In addition, in the presence of multiple SUs, the multiuser diversity gain of SUs assuming an opportunistic scheduling is also investigated. To avoid the interference at the SUs that might be caused by the random allocation scheme and obtain the maximum sum rate for SUs based on the available subcarriers, an efficient centralized sequential algorithm based on the opportunistic scheduling and random allocation (utilization) methods is proposed to ensure the orthogonality of assigned subcarriers.
\end{abstract}

\begin{IEEEkeywords}
 Random allocation,  subcarrier collision,  OFDM-based cognitive radio,  centralized scheduling, spectrum sharing, capacity, multiuser diversity.
\end{IEEEkeywords}



\section{Introduction}\label{intro}

\IEEEPARstart{A}{dvances} in communications technologies entail demands for higher data rates. One of the popular solutions to fulfill this requirement was to allocate additional bandwidth, which unfortunately is not anymore viable due to spectrum scarcity. Recent spectrum measurement campaigns, performed by agencies such as Federal Communications Commission (FCC),  reported  that the radio-frequency (RF) spectrum is being used inefficiently (see e.g.,~\cite{haykin2005cognitive} and references cited therein). Therefore, the idea of cognitive radios (CRs) was advanced as a promising approach for the efficient utilization of RF spectrum~\cite{mitola2000cognitive}.
Generally, in CR networks the usage of spectrum by cognitive (secondary) users is  maintained
by three approaches. In {\it interweave} cognitive networks, primary and secondary users are not allowed to operate simultaneously,  i.e., the secondary user (SU) accesses the spectrum while the primary user (PU) is idle. In {\it underlay} cognitive (spectrum sharing) networks, PUs  are allocated a higher priority to use the spectrum than SUs, and the coexistence of
primary and secondary users is allowed under the PU's predefined interference constraint~\cite{zhang2009ergodic, jand2011performance, suraweera2010capacity,  ban2009multi, zhang2009onpeak,  musavian2007ergodic, kang2009optimal, zhang2010investigation}, also termed  {\it interference temperature}. In {\it overlay} cognitive networks, SUs and PUs are allowed to transmit concurrently with the help of advanced coding techniques~\cite{hong2011throughput}.

One of the most challenging issues in the implementation of CR networks is to know whether at a certain physical location and moment of time  the RF spectrum is  occupied by PU(s), i.e., if there is a  sensing mechanism in place for the available spectrum~\cite{tian2007compressed, duan2010cooperative}. The challenge in deploying such a spectrum sensing mechanism is due to the uncertainties ranging from channel randomness at device
and network-level uncertainties, to  the  hidden PU problem and sensing duration~\cite{ghasemi2008spectrum, cabric2004implementation}.
There have been numerous studies to deal with these issues.
In~\cite{yucek2009survey} and references therein, a compact survey of the spectrum sensing algorithms and CR applications along with the design and implementation challenges  are  classified properly.

To understand the performance limits of a spectrum sharing system, SU capacity is a very useful performance measure. The ergodic and outage capacities of CR spectrum sharing systems in Rayleigh fading environments are studied in~\cite{musavian2007ergodic}, and a comprehensive analysis  considering various combinations of power constraints under different types of channel fading models  is performed in~\cite{kang2009optimal}.  In~\cite{ghasemi2007fundamental}, considering a point-to-point  communication scenario, the expressions for the average capacity of a single SU assuming the existence of a single PU and no PU's interference are derived for different channel fading models  such as Rayleigh, Nakagami-$m$ and Log-normal. As an extension of~\cite{ghasemi2007fundamental}, in~\cite{suraweera2010capacity}, the SU capacity assuming  PU's interference with imperfect channel knowledge, and the average bit error rate over Rayleigh channel fading were derived.  The ergodic sum capacity of CRs (SUs) with multiple access and broadcast fading channels with long-term average and short-term power constraints was established using optimal power allocation schemes in~\cite{zhang2009ergodic}. Opportunistic SUs scheduling yields multiuser diversity gain due to the channel fading  randomness, and it has been well  studied in conventional wireless systems~\cite{tse2005fundamentals, viswanath2002opportunistic}. The multiuser diversity analysis was conducted for spectrum sharing systems in~\cite{ban2009multi, zhang2010investigation}, and interweave CR networks (see~\cite{hong2011throughput} and the references cited therein).

In orthogonal frequency-division multiplexing (OFDM)-based systems, frequency spectrum is a precious and scarce resource that is divided into non-overlapping  bands, called subcarriers, and which are assigned to different cells and/or users. Starting with the early deployment of cellular mobile communication networks, efficient sharing of the available radio spectrum among the users has represented an important design problem. In conventional OFDM(A)-based systems,   stochastic subcarrier collision models have been proposed to investigate the performance of various scheduling and  deployment methods, and to assess the  inter-cell-interference (ICI) for cell-edge users~\cite{chen2011random, elayoubi2008performance,  bosisio2008interference}.

Considering the challenges and implementation issues in  CR networks in terms of spectrum sensing and subcarriers scheduling, the existing studies motivate us to investigate the performance of a  primitive (basic)  OFDM-based CR system in which the SUs randomly (blindly) utilize the available subcarriers  assuming that some of the subcarriers are utilized by the PUs. This paper focuses on such a communication scenario that assumes {\it random allocation} and {\it no spectrum sensing}.
 An immediate challenge to be addressed is the fact that the SU's subcarriers collide with PUs' subcarriers.  However,  there are no studies available to assess the effect of subcarrier collisions in such CR spectrum sharing systems.
Therefore, the requirement for a more comprehensive system analysis including the development of a stochastic model to capture the subcarrier collisions and protection of the operation of PUs in an OFDM-based CR spectrum sharing  system turns out to be crucial.
In the presence of {\it multiple} SUs, due to the random subcarrier allocation scheme,  collisions will  occur among the subcarriers used by the SUs in addition to the collisions with the subcarriers used by the PUs. The collisions among the SUs' subcarriers will decrease the system performance drastically. To overcome this issue, this paper presents also an efficient centralized algorithm that sequentially assigns the randomly selected subcarrier sets to the SUs while maintaining the orthogonality among these sets, to avoid collisions between their subcarriers. In the proposed centralized algorithm, the opportunistic scheduling of users, which yields multiuser diversity gain,  is employed and the performance limits of the system in terms of multiuser diversity gain and sum capacity of multiple SUs are studied.


The main results of this paper are next summarized:
\begin{itemize}
\item A random subcarrier allocation method, where an arbitrary $m$th SU randomly utilizes $F_m^S$ subcarriers from an available set of $F$ subcarriers in the primary network,  in an OFDM-based system is proposed. In the proposed scheme, the SUs do not have knowledge about the  PUs' subcarriers utilization, i.e., no spectrum sensing is performed. Therefore, with some probability  collisions between the subcarrier sets of PUs and SU occur. It is shown that the subcarrier collision model follows a multivariate hypergeometric distribution.

\item  Considering the average capacity as performance measure, the SU average capacity expressions under the interference constraint of PUs in the case of single or multiple PU(s) are derived. Upper and lower bounds on average capacity are derived. It is found that the average capacity of the $m$th SU scales with respect to the number of subcarriers in the sets $F$, $F_n^P$ and $F_m^S$ as\footnote{Where $F$ stands for the total number of available subcarriers in the primary network, and $F_n^P$ and $F_m^S$ are the number of subcarriers of  the $n$th PU and the $m$th SU, respectively. The notation $\Theta (\cdot)$ is introduced in {\it Definition}~\ref{sec:def_2}.} $\Theta  \left( 1 +  1/F \right )$,  $\Theta  \left( 1 - F_n^P \right )$ and $\Theta  \left( F_m^S \right )$, respectively. Furthermore, the convergence rate of average capacity as $F$ goes to infinity is found to be logarithmic.

\item To find the probability density function (PDF) and outage probability (cumulative distribution function (CDF)) of the SU capacity, which is the sum capacities of subcarriers with ``interference" and ``no-interference" from PU(s), the characteristic function (CF) and moment generation function (MGF) approaches are in general  used to obtain the PDF and CDF of sum of variates~\cite{alouini2001sum}. However, the obtained PDF and CDF for the capacity of the $i$th subcarrier for ``interference" and ``no-interference" cases are too complicated and intractable using the aforementioned approaches. Therefore, by using the moment matching method, the PDF and CDF  of the $i$th subcarrier capacity are approximated by a more tractable distribution, namely the Gamma distribution. There are various reasons for using the Gamma approximation such as being a Type-III Pearson distribution, widely used in fitting positive random variables (RVs), and its skewness and tail  are determined by its mean and variance~\cite{springer1979algebra, al2010approximation, wagner2008balance}. Even though the Gamma distribution approximation makes the analysis much easier to track the sum of capacities of all collided and collision-free subcarriers, we end up with a sum of Gamma variates with  some of the shape and scale parameters equal or non-equal, and not necessarily integer-valued. This constraint stems from the fact that individual PUs can have distinct or the same transmit power for their subcarriers.  In such a case, there are no closed-form expressions for the PDF and CDF of SU capacity. Fortunately, Moschopoulos~\cite{moschopoulos1985distribution} in 1985, proposed  a single Gamma series representation for a sum of Gamma RVs with the scale and shape parameters having the properties mentioned above. Utilizing this nice feature of Moschopoulos PDF, the PDF and CDF of SU capacity are obtained.

\item Using extreme value theory, the asymptotic analysis of multiuser diversity is investigated. The analysis conducted at this stage reveals a novel result: the limiting CDF distribution of the maximum of $\mathcal{R}$ RVs following a common Moschopoulos PDF and CDF converges to a Gumbel-type extreme value distribution as $\mathcal{R}$ converges to infinity.

\item A centralized sequential algorithm based on random allocation (utilization) and assuming an opportunistic scheduling method is proposed for scheduling the subcarriers of multiple SUs while maintaining their orthogonality. The probability mass function (PMF) of the number of subcarrier collisions for the $m$th scheduled SU in the algorithm is derived. In addition, the proposed algorithm is compared with the case, where the SUs are selected arbitrarily, i.e., no multiuser diversity gain is exploited. Last but not least, to present the impact of collisions among  the SUs' subcarriers on the sum capacity of SUs,  simulation results are provided and  compared  with the  centralized algorithm performance with and without opportunistic scheduling.

\end{itemize}


The rest of the paper is structured as follows. In Section~\ref{sec:math_def}, some essential mathematical  preliminaries and definitions are provided. The system model is presented  in Section~\ref{sec:sys_model}.
The SU capacity analysis over  arbitrary and Rayleigh fading channels is  investigated in Section~\ref{sec:average_cap}.
The multiuser diversity gain in the opportunistic scheduling of SUs  is studied in Section~\ref{sec:multiuser}.
Section~\ref{sec:scheduling} presents a centralized algorithm for orthogonal subcarrier scheduling of SUs.
The numerical and simulation results are given  in Section~\ref{sec:num_res}. Finally,  concluding remarks are drawn in Section~\ref{sec:conc}.


\section{Mathematical Preliminaries and  Definitions}\label{sec:math_def}

In this section, the hypergeometric distribution and some important definitions that are frequently used throughout the paper are provided.

\begin{proposition}[PMF of Number of Subcarrier Collisions]
When the $m$th SU randomly utilizes (allocates) $F_m^S$ subcarriers from a set of $F$ available subcarriers without replacement, and $F_n^P$ subcarriers are being used by the $n$th PU, then the PMF of the number of subcarrier collisions, $k_{nm}$, follows the hypergeometric distribution, $k_{nm}\sim \texttt{HYPG}(F_m^S,~ F_n^P, ~F)$, and is expressed as:
\begin{equation*}
\begin{split}
\textrm{Pr}(K_{nm}=k_{nm})=p(k_{nm})= \hspace*{-0.5mm}\binom{F}{F_m^S}^{\hspace*{-1mm}-1}\hspace*{-1mm}\binom{F_n^P}{k_{nm}}\hspace*{-0.5mm}\binom{
F - F_n^P}{
F_m^S - k_{nm}}\hspace*{-0.3mm},
\end{split}
\end{equation*}
where the notation  $\binom{\cdot}{\cdot}$ stands for the  binomial coefficient.

The average  number of subcarrier collisions is
\begin{equation*}
\mathbb{E}\left[k_{nm}\right] = \frac{F_m^S F_n^P }{F},
\end{equation*}
where $\mathbb{E}\left[\cdot\right]$ denotes the expectation operator.
\end{proposition}
\begin{IEEEproof} The proof can be readily shown by interpreting the process of allocating the subcarriers as selecting balls from an urn without replacement. Furthermore, the expected value of the number of subcarriers is obtained from $\mathbb{E}\left[k_{nm}\right]=  \sum_{k_{nm}}k_{nm} p(k_{nm})$.
\end{IEEEproof}

In the case of multiple PUs, the $m$th SU  might  have subcarrier collisions with up to $N$ PUs.
Let $\mathbf{k}_m=\left[k_{1m}, k_{2m}, \dots, k_{Nm}, k_{fm}\right]^T \in \mathbb{Z}^{N+1}_{0+}$ represent the number of collisions of the $m$th SU with $N$ PUs and with the collision-free subcarriers, $k_{fm}$. Then, the (joint) PMF of $\mathbf{k}_m$ is given by
\begin{equation}\label{eq:m-hypg}
\begin{split}
&  \textrm{Pr}(\mathbf{K}_m=\mathbf{k}_m)=p(\mathbf{k}_m)=\\
& \binom{F_1^P}{k_{1m}}\binom{F_2^P}{k_{2m}} \cdots  \binom{F_N^P}{k_{Nm}} \binom{F - \sum_{n=1}^{N}F_n^P}{k_{fm}}\binom{F}{F_m^S}^{-1} \\
&= \binom{F_f}{k_{fm}} \binom{F}{F_m^S}^{-1} \prod\limits_{n=1}^{N}\binom{F_n^P}{k_{nm}},
\end{split}
\end{equation}
where $F_f=F - \sum_{n=1}^{N}F_n^P$ stands for  the number of free subcarriers in the primary network. One can observe that  $\mathbf{k}_m$  follows a modified multivariate hypergeometric distribution
$\mathbf{k}_m\sim \texttt{M-HYPG}\left(F_m^S,~ \mathbf{F^P}, ~F\right)$, where $\mathbf{F^P} = \left[F_1^P, F_2^P, \dots, F_N^P, F_f\right]^T \in \mathbb{Z}^{N+1}_{0+}$, and the {\it support} of $\mathbf{k}_m$  is  given by:
\begin{equation*}
\begin{split}
& \Bigg \{ \mathbf{k}_m :  \sum_{n=1}^{N}k_{nm} + k_{fm} = F_m^S\\
& ~~ \text{and}~ k_{nm}\in \left [ \left ( F_m^S + F_n^P -F \right )^{+} , \dots,\min\left \{F_m^S, F_n^P \right \} \right ]  \Bigg \},
\end{split}
\end{equation*}
where $(x)^+=\max\{0,x\}$.

\begin{definition}[Rate of Convergence~\cite{tuyl1994acceleration}]\label{sec:def_1}
An infinite sequence $\{A_n\}$ converging to the limit $A$ is said to be {\it logarithmically} convergent if
\begin{equation*}
\lim_{n \to \infty }\frac{\left | \Delta A_{n+1}  \right |}{\left | \Delta A_{n}   \right |}\quad \text{and} \quad \lim_{n \to \infty }\frac{\left | A_{n+1} -A\right |}{\left | A_{n} -A   \right |},
\end{equation*}
both exist and are equal to unity, where $\Delta A_{n} = A_{n+1} - A_{n}$. If only $\lim_{n \to \infty }\left | \Delta A_{n+1}  \right |/\left | \Delta A_{n}   \right |=1$ holds, then the sequence $\{A_n\}$ converges {\it sublinearly} to $A$.
\end{definition}

\begin{definition}[Knuth's notations~\cite{Knuth}]\label{sec:def_2}
Let $f(n)$ and $g(n)$ be nonnegative functions. The notation:
\begin{itemize}
\item $f(n)=O(g(n))$ means that there exist  positive constants $c$ and $n_0$ such that $f(n)\le cg(n)$ for all $n\ge n_0$.
\item $f(n)=\Omega(g(n))$ means that there exist positive constants $c$ and $n_0$ such that $f(n)\ge cg(n)$ for all $n\ge n_0$, i.e., $g(n)=O(f(n))$.
\item $f(n)=\Theta (g(n))$ means that there exist  positive constants $c$, $c'$  and $n_0$ such that $cg(n) \le f(n)\le c'g(n)$ for all $n\ge n_0$, i.e., both $f(n)=O(g(n))$ and $f(n)=\Omega(g(n))$ hold.
\end{itemize}
\end{definition}

\begin{definition}\label{sec:def3}
The capacity of $m$th SU with $F_m^S$ subcarriers is defined as the summation of capacities for each subcarrier.  Let $S_{m,i}$ be the signal-to-interference plus noise ratio (SINR) for the $i$th subcarrier of the $m$th user, then the SU capacity is given by:\footnote{All logarithms in the following are with respect to the base $e$ unless otherwise stated.}
\begin{equation*}
C_m = \sum_{i=1}^{F_m}\log\left ( 1 + S_{m,i} \right ).
\end{equation*}
\end{definition}

\begin{figure*}[t]
\begin{center}
\includegraphics[width=0.8\textwidth]{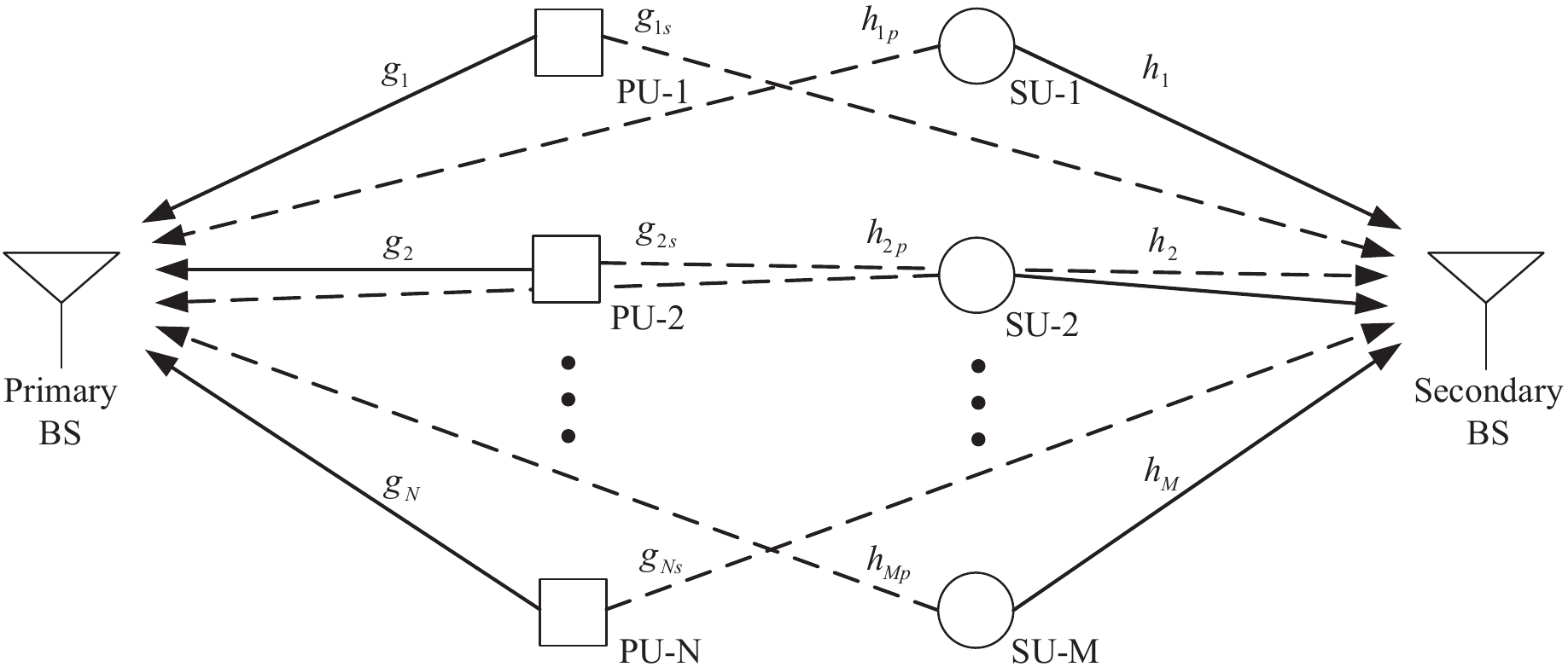}
\caption{$M$ SUs transmit to the secondary base station (SBS) using the subcarriers in the primary network with subcarrier collisions following the hypergeometric distribution for accessing PUs' subcarriers, [(- -): Interference-link (channel), (--): Desired-link (channel)].}
\label{fig:fig1}
\end{center}
\end{figure*} 

\begin{definition}[Capacity with Collisions]\label{sec:def4}
Let $S_{m,i}^{I,n}$ and $S_{m,i}^{NI}$ be the SINR for the $i$th subcarrier of the $m$th SU with ``interference" and ``no-interference" from the $n$th PU, respectively.\footnote{$S_{m,i}^{NI}$ is indeed the signal-to-noise ratio  (SNR) for the $i$th subcarrier. However, to emphasize the subcarrier collision and collision-free cases, it is called SINR with ``no-interference" from PU throughout the paper.}
If $k_{nm}$ subcarriers of the $m$th SU collide with the $n$th PU's subcarriers, then the capacity of SU in {\it Definition}~\ref{sec:def3} with subcarrier collisions can be redefined as
\begin{equation*}
C_m^1 = \sum_{i=1}^{k_{nm}}\log\left ( 1 + S_{m,i}^{I,n} \right ) + \sum_{i=1}^{k_{fm}}\log\left ( 1 + S_{m,i}^{NI} \right ),
\end{equation*}
where $k_{nm}$ and $k_{fm}=F_m^S - k_{nm}$ are hypergeometric RVs that denote the number of collided (i.e., interference) and collision-free (i.e., no-interference) subcarriers between the $n$th PU and the $m$th SU, respectively. The superscript ``1" indicates that collisions occur with only single PU's subcarriers (any arbitrary $n$th PU) in the primary network. The SU capacity expression in case of multiple $N$ PUs is given in \eqref{eq:C_m}.
\end{definition}



\section{System Model}\label{sec:sys_model}

The system model  is illustrated in Figure~\ref{fig:fig1}, where the primary and cognitive (secondary) networks consist of $N$ PUs with a primary base station (PBS) and  $M$ SUs with a secondary base station (SBS), respectively.  To preserve the quality of service (QoS) requirements of PUs in a spectrum sharing communication network, the interference power levels
caused by the SU-transmitters at the primary receiver (PBS) must
not  be larger than a predefined value ($\Psi_i,~i=1,\dots,F$) for each subcarrier, referred to
as the {\it interference temperature} (IT). It is assumed that there is no correlation among the subcarriers. Nonetheless, due to the inherent nature of random allocation (utilization) method and the high number of available subcarriers in practice, the probability of a SU to select   consecutive  subcarriers, which are practically correlated, would be considerably negligible.

The {\it channel power gains} from the $m$th SU to SBS and PBS are denoted by $h_m$ and  $h_{mp}$, respectively. Similarly,  $g_n$ and $g_{ns}$ represent the channel power gains  from the $n$th PU to PBS and SBS, respectively.
All the channel gains are assumed to be unit mean independent and identically distributed (i.i.d.) flat Rayleigh fading channels.
The channel power gains are hence exponentially distributed with unit mean.
Further, to have a tractable theoretical analysis,  it is assumed that  perfect
information  about the  interference channel power gains, $h_{mp}$, is available at SUs.
The SUs can obtain this information, referred to as channel side information
(CSI), through various ways, e.g., from the channel reciprocity condition\footnote{With the assumptions of channel reciprocity and pre-knowledge of the PBS transmit power level, SU can estimate the received signal power from PBS
when it transmits~\cite{zhang2009onpeak}.}~\cite{zhang2009onpeak, zhang2010investigation}, or from an entity called mediate band or CR network manager between the PBS and SU~\cite{ban2009multi}.
The thermal additive white Gaussian noise (AWGN) at both PUs and SUs is assumed to have circularly symmetric complex Gaussian distribution with zero mean and variance $\eta$, i.e., $\mathcal{CN}\left ( 0,\eta \right )$.
Throughout the paper, the parameters $h_{m,i}$, $h_{mp,i}$, $g_{n,i}$ and $g_{ns,i}$ denote the channel power gains associated with the $i$th subcarrier.
Furthermore, for the sake of analysis simplicity, the value of IT is assumed to be the same for all subcarriers in the system and available at the SUs, and the transmit power of each user (either PU or SU) is the same for all its subcarriers, i.e., $P_{n,i}=P_{n}$ and $P_{m,i}=P_m$.

The total number of available  subcarriers in the primary network  is denoted by $F$. The subcarrier set of each PU is assumed to be assigned by preserving the orthogonality among the sets of subcarriers for all PUs, $F_n^P$ for $n=1,\dots,N$.
SU randomly allocates the subcarriers from the available subcarriers set $F$ without having access to the information about the channel occupied  by PUs. Therefore,  SU will collide with the subcarriers of the PUs with a certain probability. Subcarrier collisions occur when SUs employ  subcarriers   which  are in use by PUs, and  the probabilistic  model for the number of subcarrier collisions follows a multivariate hypergeometric distribution.

During the evaluation of SU capacity in Section~\ref{sec:average_cap}, it is assumed that there is only {\it a single SU} (any arbitrary $m$th SU) in the cognitive network, and the collisions  occur between the subcarriers of the SU and PUs due to the random allocation scheme. This set-up can also be easily extended to multiple SUs with the assumption of no mutual interference among  SUs. However, such a framework would not be practical, since due to the random allocation method, the likelihood of  the same subcarriers being allocated to multiple SUs will be quite high. To avoid such a scenario, an efficient allocation of SUs' subcarriers is needed to preserve the orthogonality among  SUs subcarriers. Therefore, an centralized algorithm, which sequentially allocates the subcarriers to multiple SUs based on the random allocation method, while maintaining orthogonality among  SUs' subcarriers, is proposed and analyzed in Section~\ref{sec:scheduling}.


\section{Capacity of Secondary User}\label{sec:average_cap}
In this section, the average capacity of a single SU including the bounds and scaling laws with respect to the number of subcarriers for the case of an arbitrary channel fading model is investigated. Then, the Rayleigh channel fading model is used to study the impacts of the system parameters and to evaluate the expressions for the PDF and  CDF of SU capacity.


\subsection{Analysis of SU Average Capacity for General Fading}

\begin{theorem}\label{teo:avg_cap_single_PU}
The average capacity of the $m$th SU in the presence of a single ($n$th) PU is given by
\begin{equation*}
\mathbb{E}\left [C_m^1\right ]
= \frac{F_m^S}{F}\left [F_n^P\left ( \mathbb{E}\left [C_{m,i}^{I,n}\right] - \mathbb{E}\left [C_{m,i}^{NI}\right] \right ) + F\mathbb{E}\left [C_{m,i}^{NI}\right ]\right ],
\end{equation*}
where variables $C_{m,i}^{I,n}$ and $C_{m,i}^{NI}$ represent the $i$th subcarrier capacity of the $m$th SU with ``interference" and ``no-interference" from the $n$th PU, respectively.  In the case of  Rayleigh channel fading, $\mathbb{E}\left[C_{m,i}^{I,n}\right]$ and  $\mathbb{E}\left[C_{m,i}^{NI}\right]$ are given in~\eqref{eq:mean_C_I} and~\eqref{eq:mean_C_NI}, respectively.
\end{theorem}

\begin{IEEEproof}
The proof is given in Appendix~\ref{proof:teo:avg_cap_single_PU}.
\end{IEEEproof}


\begin{corollary}\label{teo:mult_PUs}
The average capacity of $m$th SU in the presence of $N$ PUs is given by
\begin{equation*}
\mathbb{E}\left[C_m\right]
=  \frac{F_m^S}{F} \left [  \sum_{n=1}^{N} F_n^P \mathbb{E} \left[ C_{m,i}^{I,n}\right] +  F_f\mathbb{E} \left[ C_{m,i}^{NI} \right] \right ].
\end{equation*}
\end{corollary}
\begin{IEEEproof}
The proof is given in Appendix~\ref{proof:teo:mult_PUs}.
\end{IEEEproof}


\subsubsection{Bounds on the Average Capacity}\label{sec:bounds}

In this section,  certain bounds on the average capacity of SU will be established. Intuitively, representing the relation of order between the average capacity of the $i$th subcarrier with PU's ``interference" and ``no-interference"  as $\mathbb{E}\left[C_{m,i}^{I,n}\right] \le \mathbb{E}\left[C_{m,i}^{NI}\right]$, the naive upper and lower bounds on the SU average capacity can be expressed as
\begin{equation}\label{eq:naive_bound1}
 F_m^S \mathbb{E}\left [ C_{m,i}^{I,n}\right]
\le \mathbb{E}\left[C_{m}^1\right]
 \le
 F_m^S \mathbb{E} \left[ C_{m,i}^{NI} \right]~,
\end{equation}
which states that the upper bound, in  the best case, is when all  SU's subcarriers are collision-free, i.e.,  all subcarriers are interference-free, $k_{fm}= F_m^S$. Similarly for the lower bound, all SU's subcarriers are colliding with the PU's subcarriers, i.e., $k_{nm}= F_m^S$.

However, the maximum and minimum  number of subcarrier collisions might not be necessarily   $F_m^S$ and $0$, respectively. The following general result holds.
\begin{corollary}
Tight upper and lower bounds on the average capacity of SU in the presence of a single PU are given by:
\begin{equation*}
\begin{split}
 k_{nm}^{\max} \mathbb{E} \left [ C_{m,i}^{I,n}\right] + k_{fm}^{\min}& \mathbb{E} \left [ C_{m,i}^{NI} \right ]
 \le \mathbb{E}\left [C_{m}^1\right] 
 \le \\
& k_{nm}^{\min} \mathbb{E} \left[ C_{m,i}^{I,n}\right]+ k_{fm}^{\max}\mathbb{E} \left[ C_{m,i}^{NI} \right],
\end{split}
\end{equation*}
where $k_{nm}^{\max}$ and $k_{nm}^{\min}$  represents the maximum and minimum number of subcarrier collisions, respectively, and are defined as $k_{nm}^{\min} = \left ( F_m^S + F_n^P -F \right )^+$ and $k_{nm}^{\max} = \min  \left \{ F_m^S , F_n^P\right \}$. Also, $k_{fm}^{\max} =  F_m^S - k_{nm}^{\min} $ and $k_{fm}^{\min} =  F_m^S - k_{nm}^{\max}$.
\end{corollary}
\begin{IEEEproof}
The number of subcarrier collisions does not depend only on SU's subcarriers but also on PU's subcarriers. Therefore, the support region of $k_{nm}$, considering the PU's  subcarriers, is  given by $\left \{ \left (F_m^S + F_n^P -F \right )^+, \dots,\min  \left \{ F_m^S , F_n^P\right \} \right\}$.
Using this support region, the bounds are established.
\end{IEEEproof}

It is worth to note that the naive upper bound, given in~\eqref{eq:naive_bound1}, on the average capacity is the limit point of capacity as the number of available subcarriers $F$ goes to infinity. Formally,
\begin{equation*}
\lim_{F \to \infty }\mathbb{E}\left[C_{m}^1\right] = F_m^S \mathbb{E}\left [ C_{m,i}^{NI} \right],
\end{equation*}
which states that for a fixed number of PU's subcarriers as the number of available subcarriers increases,  the average capacity converges to the case where no SU's  subcarrier collides.


\subsubsection{Scaling Laws for the Average Capacity}
\begin{corollary}\label{cor:scaling}
The average capacity of the $m$th SU in the presence of a single PU scales with respect to the number of subcarriers $F$, $F_m^S$ and $F_n^P$ as $\Theta  \left( 1 +  1/F \right )$,  $\Theta  \left( F_m^S \right )$ and  $\Theta  \left( 1 - F_n^P \right )$, respectively.
\end{corollary}

\begin{IEEEproof}
Using the Knuth's notation from {\it Definition}~\ref{sec:def_2}, one can infer that
\begin{equation*}\label{eq:scaling_laws}
\begin{split}
&  \lim_{F \to \infty } \frac{ \mathbb{E} \left[C_m^1\right]}{1 + \frac{1}{F}} =  \\
&  \lim_{F \to \infty }\frac{\frac{F_m^S F_n^P}{F}\left (\mathbb{E} \left[ C_{m,i}^{I,n}\right]- \mathbb{E} \left[ C_{m,i}^{NI}\right]\right) + F_m^S\mathbb{E} \left[ C_{m,i}^{NI}\right]}{1 + \frac{1}{F}} =
\\
& ~   F_m^S\mathbb{E} \left[ C_{m,i}^{NI}\right] >0. 
\end{split}
\end{equation*}

Following the same approach, one can establish the scaling laws of SU average capacity with respect to $F_m^S$ and $F_n^P$.
\end{IEEEproof}

Further, it can be also shown that for the multiple PUs case,  the average capacity of the $m$th SU is converging to the lower bound on average capacity for the single PU case as $N,F \to \infty$.
Assume without loss of generality that an infinite number of subcarriers $F$ is available.
Because the orthogonality of PUs' subcarriers is maintained, then $\sum_{n=1}^{N}F_n^P \approx F $ as $F, N \to \infty$.
Hence,
\begin{equation*}
\begin{split}
& \lim_{N, F \to \infty } \mathbb{E} \left[C_m^1\right]= \\
& \lim_{N, F \to \infty } \hspace{-1mm} \frac{F_m^S}{F} \left [  \sum_{n=1}^{N} F_n^P \mathbb{E} \left[ C_{m,i}^{I,n}\right] \hspace{-1mm} + \hspace{-1mm}  F_f\mathbb{E} \left[ C_{m,i}^{NI} \right] \right ]\hspace{-1mm}=F_m^S \mathbb{E} \left[  C_{m,i}^{I,n} \right]\hspace{-1mm} ,
\end{split}
\end{equation*}
where it is assumed that all the PUs have the same transmit power. Thus,  $\mathbb{E} \left [ C_{m,i}^{I,n} \right]$ is the same for all $N$ PUs.



\begin{corollary}\label{theo:rate_conv}
The average capacity of the $m$th secondary user in the presence of a single PU {\it converges logarithmically} to $F_m^S \mathbb{E} \left[ C_{m,i}^{NI} \right]$ as $F$ increases towards infinity:
\begin{equation}\label{eq:rate_conv}
 \mathbb{E} \left [C_m^1\right] \xrightarrow[\mathrm{with}~\log(F)]{F \to \infty}  F_m^S~\mathbb{E}\left[C_{m,i}^{NI} \right].
\end{equation}
\end{corollary}
\begin{IEEEproof}
The proof is given in Appendix~\ref{proof:theo:rate_conv}.
\end{IEEEproof}

Using similar steps, one can readily obtain the bounds and the scaling laws of the SU average capacity in the presence of multiple ($N$) PUs in the primary network.


\subsection{SU Capacity Analysis over Rayleigh Channel Fading}\label{sec:diff_fadings}
In this section, the SU capacity over a Rayleigh  channel fading model is investigated. Thus far, the CR  capacity studies in the literature  have mostly assumed two types of PUs' interference constraints on the SU transmit power: the  {\it peak power interference constraint} and the {\it average interference constraint}~\cite{suraweera2010capacity, zhang2009onpeak}.
The peak power interference constraint is adapted in this work, and an adaptive
scheme is used to adjust the transmit power of SU to maintain the QoS of PUs. Hence, the transmit power of the $m$th SU corresponding to the $i$th subcarrier is given by\footnote{Notice that due to
the random allocation, the SU transmit power is adapted (regulated) considering the worst case
scenario, as if all the subcarriers in the primary network are utilized by PUs. This condition assures the QoS requirements of PUs.}
\begin{equation*}
\begin{split}
P_{m,i}^T  = & 
\begin{cases}
P_{m,i}~, &  \Psi _{i}\geq P_{m,i}h_{mp,i} \\
\frac{\Psi _{i}}{h_{mp,i}}, &  \Psi _{i}< P_{m,i}h_{mp,i}
\end{cases}\\
=&  \min\left \{ P_{m,i}, \frac{\Psi _{i}}{h_{mp,i}} \right \}, 
\end{split}
\end{equation*}
for $ i=1,\dots , F$.

Let $\lambda_{m,i} = h_{m,i} P_{m,i}^T$, then the received SINR of the $m$th SU's  $i$th subcarrier is
\begin{equation}\label{eq:S_{m,i}^{I}_2}
S_{m,i}^{I,n} = \frac{\lambda_{m,i}}{I_{n,i}^P+\eta},\quad \text{for}~ n=1,\dots, N,
\end{equation}
where $I_{n,i}^P= P_{n,i}g_{ns,i}$ stands for the mutual interference caused by $n$th PU on the $i$th subcarrier.
In~\eqref{eq:S_{m,i}^{I}_2}, $S_{m,i}^{I,n}$ represents the SINR in case when subcarrier collision occurs.
Therefore, when there is no collision, i.e., the subcarrier is not being used by two users, there is no interference caused by PUs. Hence,
$S_{m,i}^{NI} = \lambda_{m,i}/\eta$.

The CDF of $\lambda_{m,i}$ can be obtained as follows~\cite{ji2010capacity}:
\begin{equation*}
\begin{split}
F_{\lambda_{m,i}}(x) &  =     F_{h_{mp,i}}\left ( \frac{\Psi_i }{P_{m,i}} \right) F_{\vartheta_1}( x) \\
& ~~~ +F_{\vartheta_2  |h_{mp,i}> \frac{\Psi_i }{P_{m,i}}}\left ( x ~ \Big | ~h_{mp,i}> \frac{\Psi_i }{P_{m,i}}\right),
\end{split}
\end{equation*}
where $\vartheta_1 = h_{m,i}P_{m,i}$ and $\vartheta_2 =  \Psi_i   h_{m,i}/h_{mp,i}$, with their corresponding PDFs given by $f_{\vartheta_1}(x) = e^{-x /\Psi_i}/\Psi_i$, and $f_{\vartheta_2}(x) = \Psi_i /(x+\Psi_i)^2$, respectively.
Hence the CDF and the PDF can be expressed, respectively, as
\begin{equation}\label{eq:F_{lambda_{m,i}}(x)}
\begin{split}
F_{\lambda_{m,i}}(x) & =  \left ( 1- e^{-\frac{\Psi _{i}}{P_{m,i}}} \right )\left ( 1- e^{-\frac{x}{P_{m,i}}} \right ) 
\\
& ~~~ + e^{-\frac{\Psi _{i}}{P_{m,i}}} - \frac{\Psi _{i}}{P_{m,i}+x}e^{-\frac{x+\Psi _{i}}{P_{m,i}}} \\
 & =   1- e^{-\frac{x}{P_{m,i}}}  +  \frac{x}{\Psi _{i}+x}e^{-\frac{x+\Psi _{i}}{P_{m,i}}},
\end{split}
\end{equation}
\begin{equation}\label{eq:F_{lambda_{m,i}}(x)_pdf}
\begin{split}
f_{\lambda_{m,i}}  (x) &  = \frac{d F_{\lambda_{m,i}}(x) }{dx}\\
 & =   \frac{e^{-\frac{x}{P_{m,i}}}}{P_{m,i}}\left [ 1- e^{-\frac{\Psi _{i}}{P_{m,i}}} \left ( \frac{x^2+ \Psi _{i}x - \Psi _{i}P_{m,i}}{(\Psi _{i}+x)^2} \right ) \right ].
\end{split}
\end{equation}

Similarly, by using a transformation of RVs, the PDF of $S_{m,i}^{I,n} $ with  $f_{I_{n,i}^P}(y)= e^{-y/P_{n,i}}/P_{n,i}$ can be expressed as~\cite{suraweera2010capacity}
\begin{equation}\label{eq:F_{S_{m,i}^I}(x)}
\begin{split}
F_{S_{m,i}^{I,n}}(x) &  =   \mathrm{Pr}\left ( \lambda_{m,i}< x\left ( I^P_{n,i} +\eta \right ) \right ) \\
 & = \int\limits_{0}^{\infty} F_{\lambda_{m,i}}\left (x\left ( y +\eta \right ) \right )f_{I_{n,i}^P}(y)\mathrm{d}y.
\end{split}
\end{equation}

Plugging~\eqref{eq:F_{lambda_{m,i}}(x)} into~\eqref{eq:F_{S_{m,i}^I}(x)}, it follows that
\begin{equation*}
\begin{split}
F_{S_{m,i}^{I,n}}(x)  &  =  1- \frac{\left (1- e^{-\frac{\Psi_i}{P_{m,i}}} \right ) e^{-\frac{x \eta }{P_{m,i}}} }{1+\frac{x P_{n,i}}{P_{m,i}}} -\frac{\Psi_i}{x P_{n,i}} e^{\frac{\Psi_i}{x P_{n,i}}+\frac{\eta}{P_{n,i}}}\\
&  ~~~  \times \Gamma \left (0, \left ( \eta + \frac{\Psi_i}{x}\right )  \left ( \frac{1}{P_{n,i}} + \frac{x}{P_{m,i}}\right ) \right ),
\end{split}
\end{equation*}
where the upper incomplete Gamma function is defined as $\Gamma( x, y ) =\int_{y}^{\infty} t^{x-1} e^{-t}\mathrm{d}t $, and the derivation of CDF yields the PDF
\begin{equation}\label{eq:f_{S_{m,i}^I}(x)}
\begin{split}
&  f_{S_{m,i}^{I,n}}(x) = \\
&  \frac{x \eta P_{n,i} + P_{m,i} (\eta + P_{n,i})}{(x P_{n,i}+P_{m,i} )^2}\left ( e^\frac{\Psi_i}{P_{m,i}} -1 \right )e^{-\frac{x \eta +\Psi_i }{P_{m,i}}} 
  + \frac{\Psi_i}{x^3 P_{n,i}^2}\\
 & \times e^{\frac{x \eta +\Psi_i }{x P_{n,i}}} 
  \left [ (\Psi_i + xP_{n,i}) \Gamma \left (0, \left ( \eta + \frac{\Psi_i}{x}\right )  \left ( \frac{1}{P_{n,i}} + \frac{x}{P_{m,i}}\right ) \hspace*{-0.5mm}\right ) 
\right.    \\
&  \left.  +
 \frac{x P_{n,i} (x^2 \eta P_{n,i} - \Psi_i P_{m,i})}{(x \eta  + \Psi_i )(x P_{n,i} + P_{m,i})}   e^{-\left ( \eta + \frac{\Psi_i}{x}\right )  \left ( \frac{1}{P_{n,i}} + \frac{x}{P_{m,i}}\right ) }  \right ].
\end{split}
\end{equation}

Similarly, when there is no primary interference using~\eqref{eq:F_{lambda_{m,i}}(x)_pdf} and the transformation $f_{S_{m,i}^{NI}}(x) = \eta f_{\lambda_{m,i}}(\eta x)$, it follows that
\begin{equation}\label{eq:f_S_{m,i}}
\begin{split}
& f_{S_{m,i}^{NI}} (x) =\\
&  \frac{\eta e^{-\frac{\eta x}{P_{m,i}}}}{P_{m,i}}\left [ 1- e^{-\frac{\Psi _{i}}{P_{m,i}}} \left ( \frac{(\eta x)^2+ \Psi _{i} \eta x - \Psi _{i}P_{m,i}}{(\Psi _{i}+ \eta x)^2} \right ) \right ],
\end{split}
\end{equation}
and the CDF is given by 
\begin{equation}\label{eq:F_S_{m,i}}
\begin{split}
F_{S_{m,i}^{NI}}(x)
&= 1- e^{-\frac{\eta x}{P_{m,i}}}  +  \frac{\eta x}{\Psi _{i}+ \eta x}e^{-\frac{\eta x+\Psi _{i}}{P_{m,i}}}.
\end{split}
\end{equation}

Finally, the desired expressions for the PDFs of $C_{m,i}^{I,n}$ and $C_{m,i}^{NI}$ can be obtained by transforming the RVs as follows:
\begin{equation}\label{eq:cap_eq}
\begin{split}
f_{C_{m,i}^{I,n} }(x)& = \left | \frac{dy}{dx} \right | f_{S_{m,i}^{I,n}}(y) \bigg |_{y=e^{x}-1}  = e^x f_{S_{m,i}^{I,n}}(e^{x}-1),\\
f_{C_{m,i}^{NI} }(x)&   = e^x f_{S_{m,i}^{NI}}(e^{x}-1).
\end{split}
\end{equation}

Using {\it Definition}~\ref{sec:def4}, for any arbitrary $m$th SU and multiple ($N$) interfering PUs, the instantaneous SU capacity with subcarrier collisions is given by
\begin{equation}\label{eq:C_m}
\begin{split}
C_m &  =  \sum_{i=1}^{k_{1m}}\underbrace{\log\left ( 1 + S_{m,i}^{I,1} \right )}_{C_{m,i}^{I,1}}  +
\cdots + \sum_{i=1}^{k_{Nm}}\underbrace{\log\left ( 1 + S_{m,i}^{I,N} \right )}_{C_{m,i}^{I,N}}
 \\ 
& ~~~+ \sum_{i=1}^{k_{fm}}\underbrace{\log\left ( 1 + S_{m,i}^{NI} \right )}_{C_{m,i}^{NI}} \\
& =  \underbrace{\sum_{n=1}^{N} \overbrace{\sum_{i=1}^{k_{nm}}{C_{m,i}^{I,n}} }^{C_{m}^{I,n}}}_{C_m^I}
+\underbrace{ \sum_{i=1}^{k_{fm}}C_{m,i}^{NI} }_{C_{m}^{NI}}.
\end{split}
\end{equation}

There are two types of well known methods available to evaluate the distribution for sum of variates, namely, the characteristic function (CF) and the moment generating function (MGF) based methods~\cite{alouini2001sum}. Unfortunately, by employing these methods, it is often hard and intractable to obtain explicit closed form expressions for the PDF and  CDF of SU capacity in~\eqref{eq:C_m} from~\eqref{eq:f_{S_{m,i}^I}(x)}-\eqref{eq:cap_eq}. Even if we obtain, it will hardly provide any insights because of the complicated expressions. Therefore, in order to sum up the rates for the cases of interference and no-interference, we will approximate the PDFs of $C_{m,i}^{I,n}$ and $C_{m,i}^{NI}$ using a Gamma distribution. There are important properties of the Gamma distribution that are suitable for approximating the PDFs of the variables $C_{m,i}^{I,n}$ and $C_{m,i}^{NI}$. First, the sum of Gamma distributed RVs with the same scale parameters is another Gamma distributed RVs. Second, the skewness and tail of  distribution are similar for the whole range of interest and are determined by  mean and variance~\cite{wagner2008balance}. Last but not least, Gamma distribution is a Type-III Pearson distribution which is widely used in fitting {\it positive} RVs~\cite{springer1979algebra, al2010approximation, wagner2008balance}.
In addition, since Gamma distribution is uniquely determined by its mean and variance,
we employed the moment matching method to the first two moments: mean and variance.

\begin{definition}
$X$  follows a Gamma distribution, $X\sim \mathcal{G}(\alpha, \beta)$, if the corresponding PDF of $X$ with scale and shape parameters, $\beta > 0$ and $\alpha > 0$, respectively, is given by
\begin{equation*}
f_{X}(x)= \frac{x^{\alpha -1} \exp\left ( - \frac{x}{\beta }\right )}{\beta^\alpha \Gamma(\alpha) }U(x),
\end{equation*}
where $U(\cdot)$ denotes the unit step function, and the Gamma function is defined as $\Gamma( x ) =\int_{0}^{\infty} t^{x-1} e^{-t}\mathrm{d}t$.
\end{definition}

Since the mean and variance of Gamma distribution are $\alpha \beta$ and $\alpha \beta^2$, respectively, mapping the first two moments with the PDFs of  $ C_{m,i}^{I,n} $ and $C_{m,i}^{NI}$ yields
\begin{equation*}
\begin{split}
\alpha^I_n & = \frac{\left( \mathbb{E}\left[C_{m,i}^{I,n}\right]\right )^2}{\mathbf{var}\left[C_{m,i}^{I,n}\right]}, \quad \beta^I_n = \frac{\mathbf{var}\left [C_{m,i}^{I,n}\right]}{\mathbb{E}\left [C_{m,i}^{I,n}\right]} , \\
\alpha^{NI} & = \frac{\left ( \mathbb{E}\left [C_{m,i}^{NI}\right]\right )^2}{\mathbf{var}\left [C_{m,i}^{NI}\right]}, \quad \beta^{NI} = \frac{\mathbf{var}\left [C_{m,i}^{NI}\right]}{\mathbb{E}\left [C_{m,i}^{NI}\right]},
\end{split}
\end{equation*}
for $n=1,2,\dots, N$, and $\mathbf{var}(x)$ denotes the variance of $x$.

From~\cite{suraweera2010capacity}, using~\eqref{eq:f_{S_{m,i}^I}(x)}-\eqref{eq:cap_eq}, the average capacity of  $C_{m,i}^{I,n}$  and $C_{m,i}^{NI}$  can be expressed, respectively, as
\begin{equation}\label{eq:mean_C_I}
\begin{split}
 \mathbb{E}\left [C_{m,i}^{I,n}\right]   & =  \int\limits_{0}^{\infty}xf_{C_{m,i}^{I,n}}(x)\mathrm{d}x\\
 & = \int\limits_{0}^{\infty}\log (1+x)f_{S_{m,i}^{I,n}}(x)\mathrm{d}x  =   \frac{ 1- e^{-\frac{\Psi_i}{P_{m,i}}} }{1-\frac{P_{n,i}}{P_{m,i}}} 
\\
&\quad \times \left ( \Gamma \left ( 0,\frac{\eta}{P_{m,i}} \right ) e^{\frac{\eta}{P_{m,i}}}  - \Gamma \left ( 0,\frac{\eta}{P_{n,i}} \right ) e^{\frac{\eta}{P_{n,i}}}\right )  \\
& \quad  +  \frac{\Psi_i}{P_{n,i}}e^{\frac{ \eta }{P_{n,i}}}\int\limits_{0}^{\infty}\Gamma \left (0, \left ( \eta + \frac{\Psi_i}{x}\right )  \right. \\
& \quad \times \left. \left ( \frac{1}{P_{n,i}} + \frac{x}{P_{m,i}}\right ) \right )  \frac{e^{\frac{\Psi_i}{xP_{n,i}}}}{x(1+x)}\mathrm{d}x,
\end{split}
\end{equation}
and
\begin{equation}\label{eq:mean_C_NI}
\begin{split}
\hspace*{-3mm}\mathbb{E}\left [C_{m,i}^{NI}\right] & =  \int\limits_{0}^{\infty}xf_{C_{m,i}^{NI}}(x)\mathrm{d}x = \int\limits_{0}^{\infty}\log (1+x)f_{S_{m,i}^{NI}}(x)\mathrm{d}x \\
& =   \Gamma \left ( 0,\frac{\eta}{P_{m,i}} \right ) e^{\frac{\eta}{P_{m,i}}} \left ( 1+ \frac{e^{-\frac{\Psi_{i}}{P_{m,i}}}\eta }{\Psi_{i}-\eta} \right ) \\ &~~~ + \frac{\Psi_{i}}{\eta -\Psi_{i}}  \Gamma \left ( 0,\frac{\Psi_{i}}{P_{m,i}}  \right ).
\end{split}
\end{equation}

The variance  of $C_{m,i}^{I,n}$  is given  by
\begin{equation*}
\mathbf{var}\left [C_{m,i}^{I,n}\right]= \mathbb{E}\left [\left (C_{m,i}^{I,n}\right)^2\right]  - \left(\mathbb{E}\left [C_{m,i}^{I,n}\right]\right )^2,
\end{equation*}
where the second moment of $C_{m,i}^{I,n}$ is expressed as
\begin{equation*}
\begin{split}
\mathbb{E}\left [\left (C_{m,i}^{I,n}\right)^2\right] = & \int\limits_{0}^{\infty}\left [\log(1+x)\right]^2 f_{S_{m,i}^{I,n}}(x)\mathrm{d}x\\
= &  \int\limits_{0}^{\infty}\frac{2\log(1+x)}{1+x} \left [1-F_{S_{m,i}^{I,n}}(x)\right]\mathrm{d}x \\
\simeq & \sum_{j=1}^{N_p}w_{j}\frac{2\log(1+s_j)}{1+s_j}\left [1-F_{S_{m,i}^{I,n}}(s_j)\right],
\end{split}
\end{equation*}
where the second equality is obtained by using integration by parts~\cite{suraweera2010capacity}. The resulting integral is readily estimated by employing Gauss-Chebyshev quadrature (GCQ) formula, where the weights ($w_j$) and abscissas ($s_j$) are defined in \cite[Eqs. (22) and (23)]{yilmaz2010mgf}, respectively. The truncation index $N_p$ could be chosen to make the approximation error negligibly small such as $N_p = 50$ for a sufficiently accurate result.

\begin{figure*}[t]
  \centering
  \subfigure[]{ \label{fig:fig2_a} \includegraphics[width=0.47\textwidth]{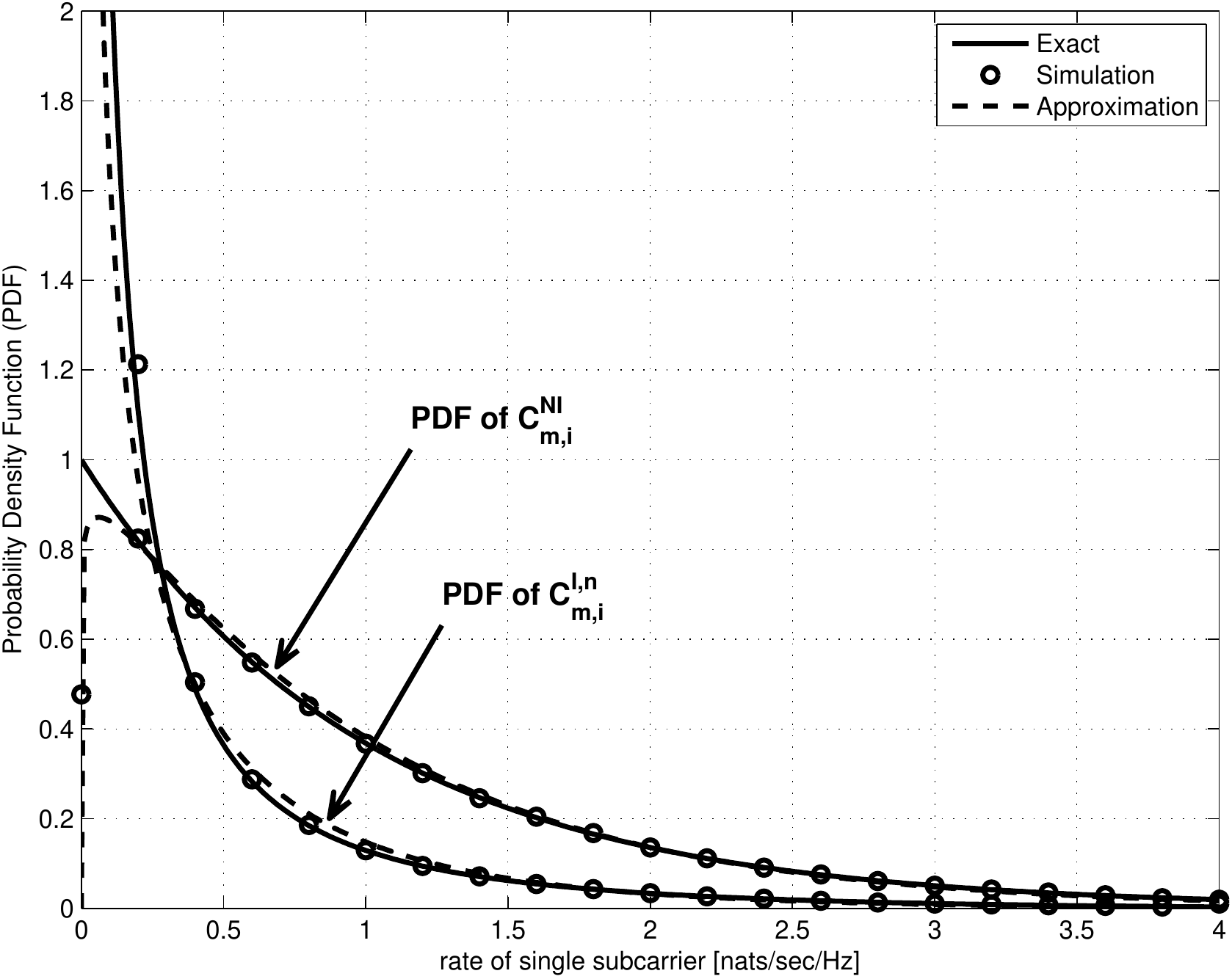}}
\hspace{3mm}
  \subfigure[]{ \label{fig:fig2_b} \includegraphics[width=0.47\textwidth]{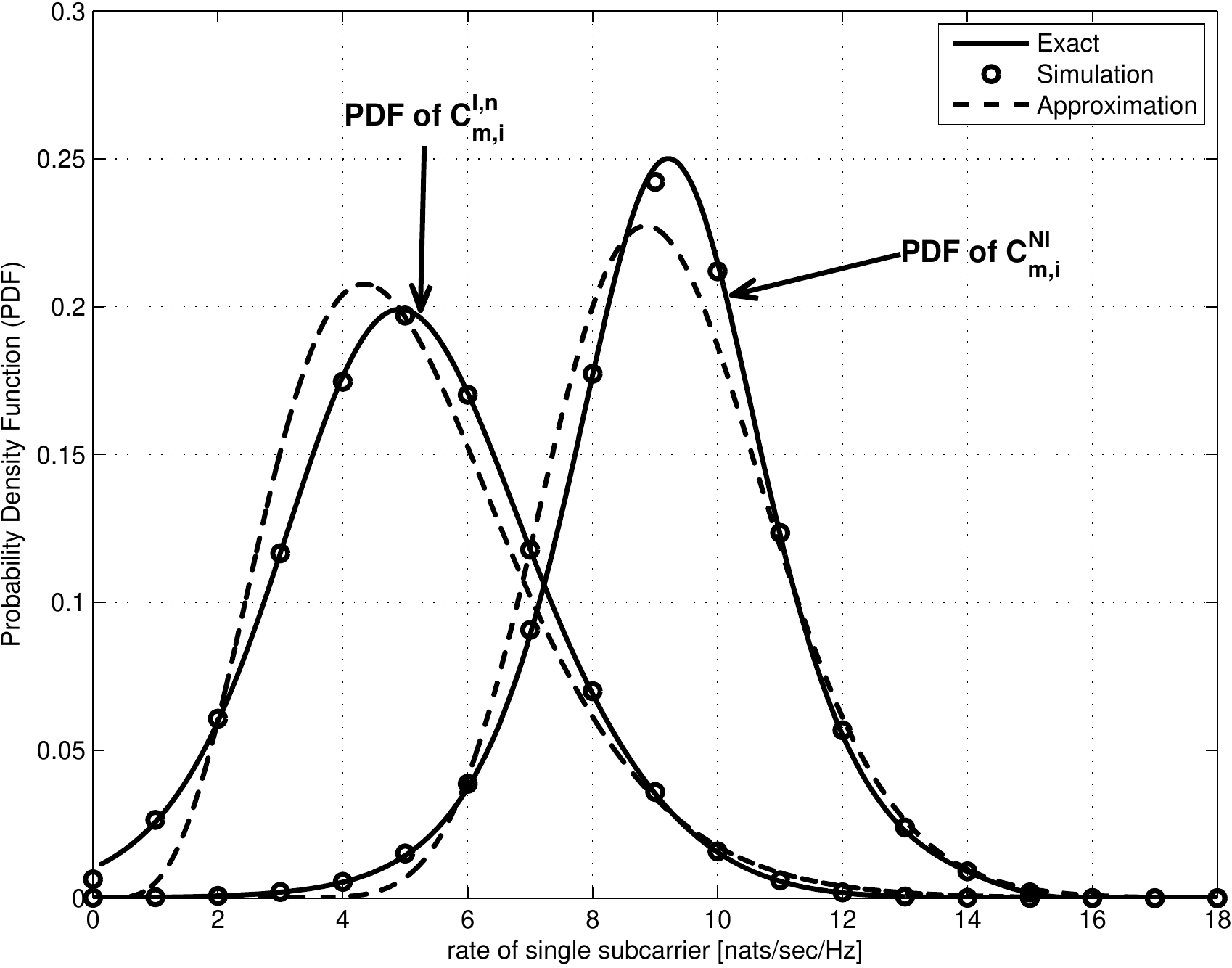}}
  \caption{Comparison between the exact and approximation of $f_{C_{m,i}^{I,n} }(x)$ and $f_{C_{m,i}^{NI} }(x)$ using the PDF of Gamma distribution for (a)  $P_{m,i}=20$ dB, $P_{n,i}=10$ dB, $\Psi_i=0$ dB and $\eta = 1$,  (b) $P_{m,i}=40$ dB, $P_{n,i}=0$ dB, $\Psi_i=20$ dB and $\eta = 0.01$.}
  \label{fig:fig2}
\end{figure*}

Similarly, the variance of $C_{m,i}^{NI}$ is expressed as
\begin{equation*}
\mathbf{var}\left [C_{m,i}^{NI}\right]= \mathbb{E}\left [\left (C_{m,i}^{NI}\right)^2\right]  -\left(\mathbb{E}\left [C_{m,i}^{NI}\right]\right )^2,
\end{equation*}
where the second moment of $C_{m,i}^{NI}$ is calculated as follows
\begin{equation*}
\mathbb{E}\left [\left (C_{m,i}^{NI}\right)^2\right] \simeq \sum_{j=1}^{N_p}w_{j}\frac{2\log(1+s_j)}{1+s_j}\left[1-F_{S_{m,i}^{NI}}(s_j)\right].
\end{equation*}

Therefore, using the Gamma approximation, the capacities are approximated as $C_{m,i}^{I,n} \sim \mathcal{G} \left (\alpha^I_n, \beta^I_n \right )$ and $C_{m,i}^{NI}\sim \mathcal{G} \left (\alpha^{NI}, \beta^{NI} \right )$.

In Figure~\ref{fig:fig2}, the exact and approximative expressions of $f_{C_{m,i}^{I,n} }(x)$ and  $f_{C_{m,i}^{NI} }(x)$, including  the simulations results, for different system parameters  are shown. It can be observed that the approximation is very close to the exact results.

Since both $C_{m,i}^{I,n}$ and $C_{m,i}^{NI}$ are i.i.d. for given $k_{nm}$, the conditional characteristic functions for  the rate sums
$ \sum_{i=1}^{k_{nm}}C_{m,i}^I$ and
$\sum_{i=1}^{k_{fm}}C_{m,i}^{NI} $ can be expressed as follows
\begin{equation*}
\begin{split}
\Phi_{C_{m}^{I,n}}(\omega | k_{nm} ) &= \left (  \Phi_{C_{m,i}^{I,n}}(\omega ) \right )^{k_{nm}} = \left ( 1- j\omega \beta^I_n \right )^{-\alpha^I_n k_{nm}},
\\
\Phi_{C_{m}^{NI}}(\omega | k_{nm} ) &= \left (  \Phi_{C_{m,i}^{NI}}(\omega ) \right )^{k_{fm}} = \left ( 1- j\omega \beta^{NI} \right )^{-\alpha^{NI}k_{fm}},
\end{split}
\end{equation*}
where $\Phi_{C_{m,i}^{I,n}}(\omega | k_{nm})$ and $\Phi_{C_{m,i}^{NI}}(\omega | k_{nm} )$ are the characteristic functions of  $f_{C_{m,i}^{I,n}}(x|k_{nm})$  and  $f_{C_{m,i}^{NI}}(x|k_{nm})$, respectively. Using the nice feature of the Gamma distribution that the sum of i.i.d. Gamma distributed RVs, with the same scale parameters ($\beta$) is another Gamma distributed RV, the conditional PDFs are expressed as follows
\begin{equation}\label{eq:f_{C_m^{I,n}}}
\begin{split}
f_{C_m^{I,n}| k_{nm}}(x | k_{nm}) &= \mathcal{G} \left (\alpha^I_n k_{nm}, \beta^I_n \right ),
\\
f_{C_m^{NI}| k_{nm}}(x | k_{nm}) & = \mathcal{G} \left (\alpha^{NI}  k_{fm}, \beta^{NI} \right ).
\end{split}
\end{equation}

In~\eqref{eq:C_m}, even though  the conditional PDFs of  $C_m^{I,n}$  and $ C_m^{NI}$ are obtained,  to find the PDF expression for $C_m$, we first need to evaluate the PDF of $C_m^{I}$, and then the PDF of  its sum with $C_m^{NI}$. At this point, one needs to be aware that  there are $N+1$ terms in~\eqref{eq:C_m}, and each follows a Gamma distribution where the shape ($\alpha$) and scale ($\beta$)  parameters  can be arbitrary. Therefore, the aforementioned feature of Gamma distribution for a sum of  Gamma variates cannot be employed here.

Expressions for the PDF of sum of Gamma RVs  are derived by  Moschopoulos~\cite{moschopoulos1985distribution}, Mathai~\cite{mathai1982storage}, and Sim~\cite{sim1992point}. In addition, constraining the shape parameters to take  integer values\footnote{If the shape parameter is an integer, Gamma distribution is referred to as Erlang distribution.} and be all distinct, by using the convolution of PDFs Coelho~\cite{coelho1998generalized} and Karagiannidis {\it et al.}~\cite{karagiannidis2006closed}, or partial-fractions methods Mathai~\cite{mathai1982storage}, derived an expression for the PDF of a sum of Gamma RVs.
Nevertheless, Moschopoulos PDF provides a mathematically tractable solution that it does not restrict the scale and shape parameters to be necessarily integer-valued  or all distinct~\cite{mathai1982storage}.
Therefore, the following theorem will help us in this regard.

\begin{theorem}[Moschopoulos, 1985~\cite{moschopoulos1985distribution}]\label{teo:moschopoulos}
Let $\left \{ X_s \right \}_{s=1}^\mathcal{S}$ be independent but not necessarily identically distributed Gamma variates with parameters $\alpha_s$ and $\beta_s$, respectively, then the PDF of $Y = \sum_{s=1}^\mathcal{S} X_s$    can be expressed as
\begin{equation}\label{eq:mos_pdf}
\begin{split}
f_Y(y) = \prod_{s=1}^{\mathcal{S}}\left ( \frac{\beta_{1}}{\beta_s} \right )^{\alpha_s} \sum_{k=0}^{\infty} \frac{\delta_k y^{\sum_{s=1}^{\mathcal{S}}\alpha_s+k-1} \exp\left ( -\frac{y}{\beta_{1}} \right )}{\beta_{1}^{\sum_{s=1}^{\mathcal{S}}\alpha_s+k}\Gamma\left ( \sum\limits_{s=1}^{\mathcal{S}} \alpha_s +k \right )}U(y),
\end{split}
\end{equation}
where $\beta_{1} = \min_s \{ \beta_s\}$, and the coefficients $\delta_k$  can be obtained recursively by the formula
\begin{equation*}
\begin{split}
& \delta_0 =1  \\
& \delta_k = \frac{1}{k+1} \sum\limits_{i=1}^{k+1}\left [ \sum\limits_{j=1}^{\mathcal{S}}\alpha_j \left ( 1- \frac{\beta_{1}}{\beta_j} \right )^i \right ]\delta_{k+1-i}  
\end{split}
\end{equation*}
for $k=0,1,2,\dots$
\end{theorem}

\begin{IEEEproof}
See~\cite{moschopoulos1985distribution}.
\end{IEEEproof}

\begin{figure*}[t]
  \centering
  \subfigure[]{ \label{fig:fig3_a} \includegraphics[width=0.47\textwidth]{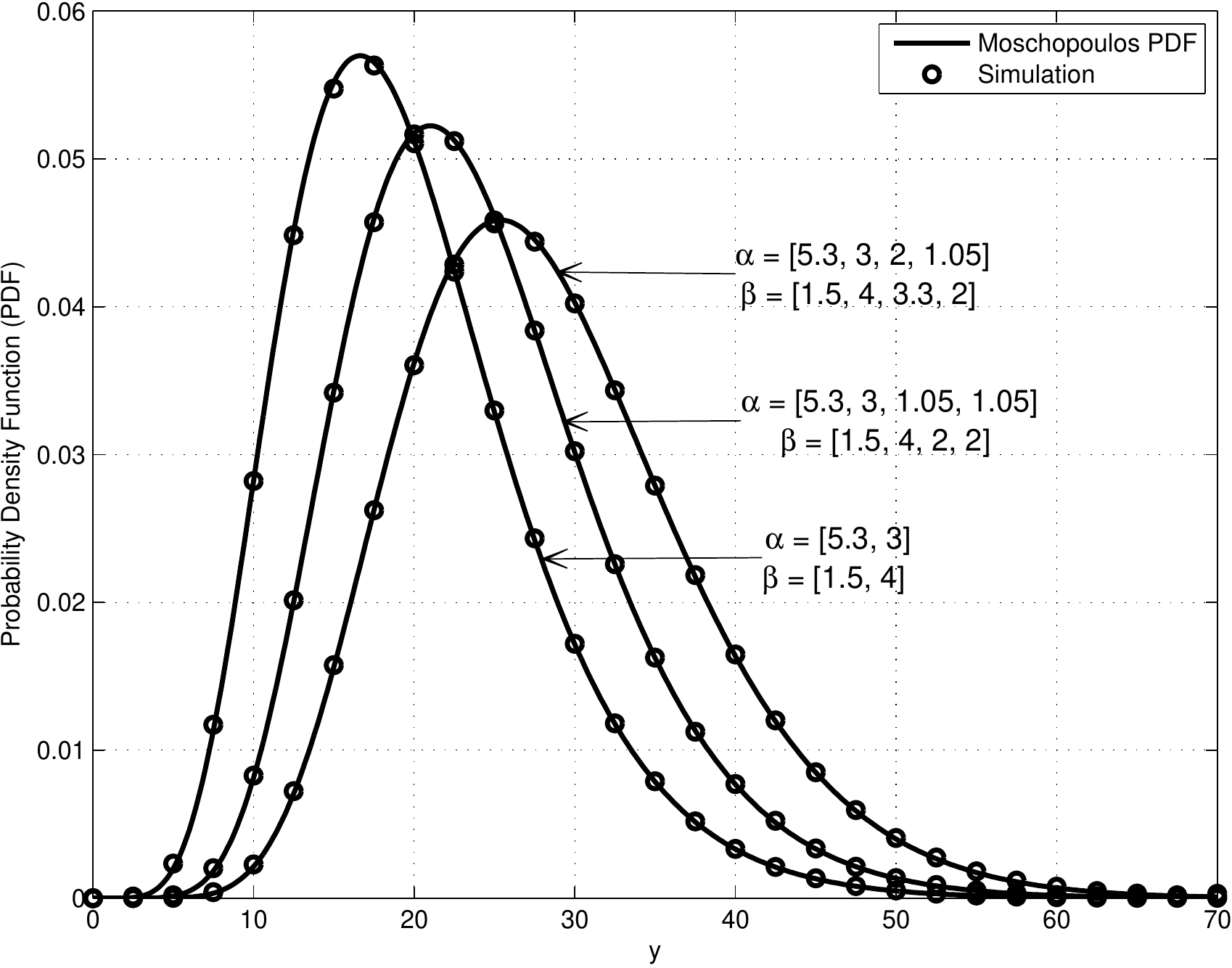}}
\hspace{3mm}
  \subfigure[]{ \label{fig:fig3_b} \includegraphics[width=0.47\textwidth]{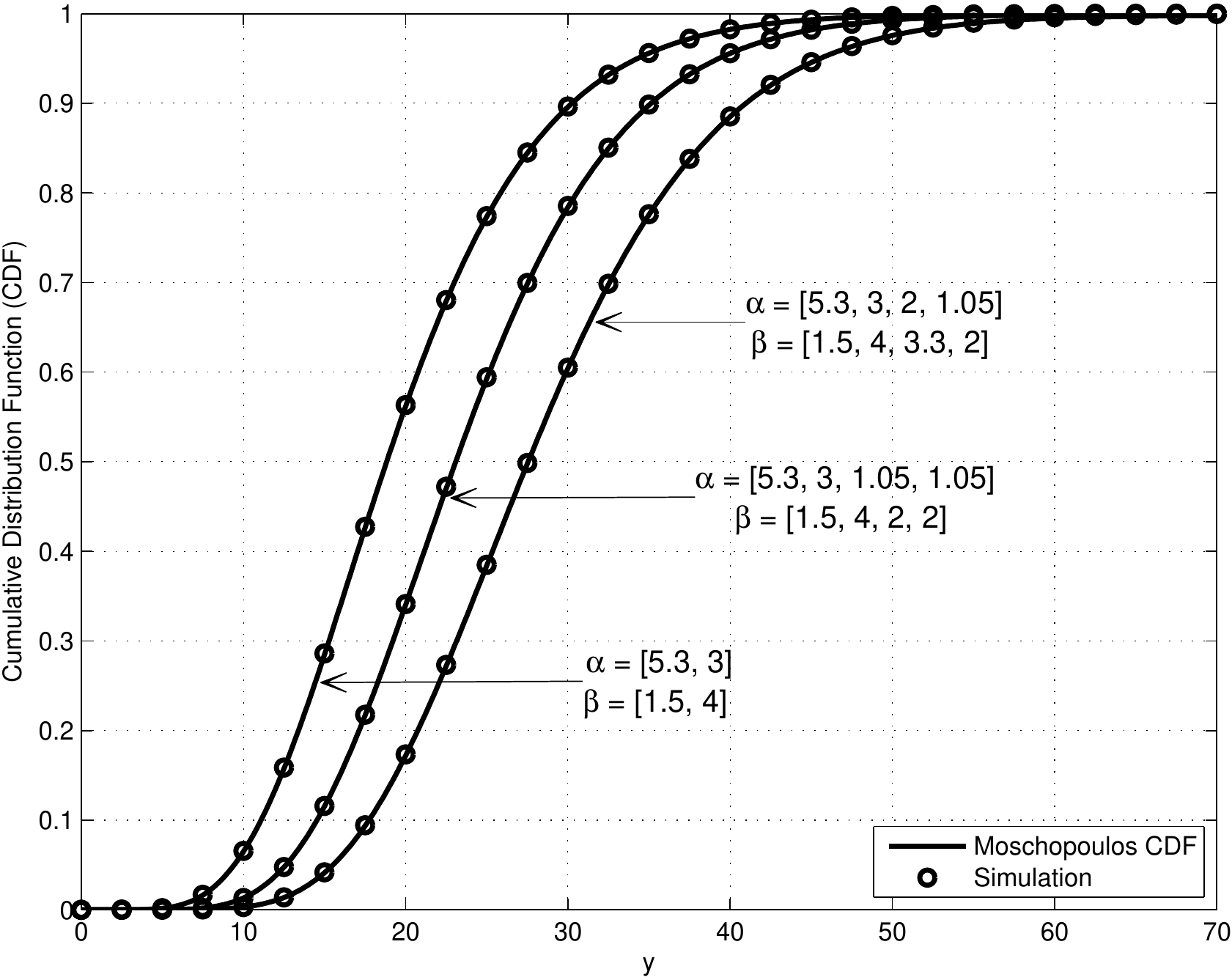}}
  \caption{Moschopoulos (a) PDF~\eqref{eq:mos_pdf} and  (b) CDF~\eqref{eq:mos_cdf2} for $h=25$, and the total number of Gamma distributed RVs in sum as $\mathcal{S}=4$ and $\mathcal{S}=2$.}
  \label{fig:fig3}
\end{figure*}

The Moschopoulos PDF provides a nice and tractable representation of sum of Gamma variates in terms of a single Gamma series with a simple recursive formula to calculate the coefficients. This representation is applicable for any arbitrary shape parameters $\{\alpha_s\}_{s=1}^{\mathcal{S}}$ and scale parameters $\{\beta_s\}_{s=1}^{\mathcal{S}}$ including the possibility of having some of  the parameters  identical.

The CDF of $Y$ can be obtained from the PDF as $F_Y(y)  = \int_{-\infty}^{y}f_Y(x)\mathrm{d}x$. Therefore,
\begin{equation}\label{eq:mos_cdf}
\begin{split}
\hspace*{-1mm}F_Y(y) =&  \prod_{s=1}^{\mathcal{S}}\left ( \frac{\beta_{1}}{\beta_s} \right )^{\alpha_s} \sum_{k=0}^{\infty} \frac{\delta_k
}{\beta_{1}^{\sum_{s=1}^{\mathcal{S}}\alpha_s+k}\Gamma\left ( \sum\limits_{s=1}^{\mathcal{S}} \alpha_s +k \right )} 
\\
& \times \int\limits_{0}^{y}  x^{\sum_{s=1}^{\mathcal{S}}\alpha_s+k-1} \exp\left ( -\frac{x}{\beta_{1}} \right ) \mathrm{d}x.
\end{split}
\end{equation}

The interchange of summation and integration above is justified using the uniform convergence of~\eqref{eq:mos_pdf} (see e.g.,~\cite{moschopoulos1985distribution} for a rigorous proof).  From~\cite{suraweera2006outage}, we can simplify~\eqref{eq:mos_cdf} by using $\int_{0}^{u}x^{\nu -1}e^{-\mu x}\mathrm{d}x = \mu^{-\nu}\gamma\left ( \nu, \mu u \right ) $ for $\Re \left [ \nu>0 \right ]$ \cite[pg. 346, Sec. 3.381, Eq. 1]{gradshtein2000table}, where $\gamma( \cdot, \cdot )$ is  the lower incomplete Gamma function and is defined as $\gamma( x, y ) =\int_{0}^{y} t^{x-1} e^{-t}dt $. Hence,
\begin{equation}\label{eq:mos_cdf2}
\begin{split}
F_Y(y) =&  \prod_{s=1}^{\mathcal{S}}\left ( \frac{\beta_{1}}{\beta_s} \right )^{\alpha_s} \sum_{k=0}^{\infty} \delta_k\frac{ \gamma\left ( \sum\limits_{s=1}^{\mathcal{S}}\alpha_s+k, \frac{y}{\beta_{1}} \right )
}{\Gamma\left ( \sum\limits_{s=1}^{\mathcal{S}} \alpha_s +k \right )}
\\
= & \prod_{s=1}^{\mathcal{S}}\left ( \frac{\beta_{1}}{\beta_s} \right )^{\alpha_s} \sum_{k=0}^{\infty} \delta_k \mathcal{P}\left ( \sum\limits_{s=1}^{\mathcal{S}}\alpha_s+k, \frac{y}{\beta_{1}}\right ),
\end{split}
\end{equation}
where  $\mathcal{P}(\cdot,\cdot)$ is the regularized (also termed  normalized) incomplete Gamma function and defined as\footnote{ For integer values of $\sum_{s=1}^{\mathcal{S}}\alpha_s+k$, using~\cite[Eq. 8.353.6]{gradshtein2000table} the regularized incomplete Gamma function can be further simplified to  $\mathcal{P}\left ( \sum_{s=1}^{\mathcal{S}}\alpha_s+k, \frac{y}{\beta_{1}}\right ) = 1- \exp({-y/\beta_{1}}) \sum_{j=1}^{\sum_{s=1}^{\mathcal{S}}\alpha_s+k-1} \frac{1}{j!}\left (\frac{y}{\beta_{1}} \right )^j$ .} $\mathcal{P}(a,z)= \frac{\gamma(a,z)}{\Gamma(a)} = 1- \frac{\Gamma(a,z)}{\Gamma(a)}$.
 For practical purposes, based on the required accuracy of application one may use the first $h$, i.e., $k=h-1$, terms in the sum series~\eqref{eq:mos_pdf}. The expression for truncation error is given in~\cite{moschopoulos1985distribution}. In Figure~\ref{fig:fig3}, the Moschopoulos PDF and CDF  are shown for $\mathcal{S}=4$ and $\mathcal{S}=2$ where  only the first 25 terms in the infinite sum series, i.e., $h=25$, are considered. One can observe that the  Moschopoulos PDF and CDF perfectly agree with the simulation results for same values of $\alpha$ and $\beta$. Since in our system model,  with some probability the  transmit power of PUs $P_{n,i}$ for $n=1,\dots,N$, can be the same, which means that the corresponding $\alpha_n^I$ and $\beta_n^I$ are the same. Such a scenario can arise when the PUs  are at the same distance from their corresponding common PBS.

Recall that from~\eqref{eq:C_m} and~\eqref{eq:f_{C_m^{I,n}}}, we have to evaluate the PDF of the sum $C_{m}^{I,1} + C_{m}^{I,2}+  \cdots + C_{m}^{I,N} +C_{m}^{NI}$,  for a  given number of set of subcarrier collisions
$\mathbf{k}_m=\left[k_{1m}, k_{2m}, \dots, k_{Nm}, k_{fm}\right]$. Recall also that
 $C_{m}^{I}$ and $C_{m}^{NI}$
 are Gamma distributed and  independent but not necessarily identical. Therefore, the conditional PDF of their sum can be expressed by means of {\it Theorem}~\ref{teo:moschopoulos} as given in~\eqref{eq:f_C_mm}, 
\begin{figure*}
\begin{equation}\label{eq:f_C_mm}
\begin{split}
f_{C_m| \mathbf{K}_m}(x | \mathbf{k}_m)  =&  \left ( \frac{\beta_{\min}}{\beta^{NI}} \right )^{\alpha^{NI}k_{fm}}\prod_{n=1}^{N}\left ( \frac{\beta_{\min}}{\beta_n^{I}} \right )^{\alpha_n^{I}k_{nm}} \sum_{k=0}^{\infty} \frac{\delta_k x^{\sum_{n=1}^{N}\alpha_n^Ik_{nm} + \alpha^{NI}k_{fm}+k-1} \exp\left ( -\frac{x}{\beta_{\min}} \right ) U(x) }{\beta_{\min}^{\sum_{n=1}^{N}\alpha_n^Ik_{nm} + \alpha^{NI}k_{fm}+k} ~\Gamma\left ( {\sum\limits_{n=1}^{N}\alpha_n^Ik_{nm} + \alpha^{NI}k_{fm}+k}  \right )},
\end{split}
\end{equation}
\vspace*{0.3cm}
\\ 
\rule[0.2cm]{1\textwidth}{0.017cm} 
\end{figure*}
where  $\beta_{\min} = \min \{ \beta^{I}_1, \beta^{I}_2, \dots, \beta^{I}_N, \beta^{NI}\}$, and the coefficients $\delta_k$ are obtained recursively as follows:
\begin{equation*}
\begin{split}
& \delta_0 =1  \\
& \delta_k = \frac{1}{k+1} \sum\limits_{i=1}^{k+1}\left [ \sum\limits_{j=1}^{N}\alpha_i^I k_{jm} \left ( 1- \frac{\beta_{\min}}{\beta_j^I} \right )^i \right.
\\& \left. ~~~~~ + \alpha^{NI} k_{fm} \left ( 1- \frac{\beta_{\min}}{\beta^{NI}} \right )^i\right ]\delta_{k+1-i} ~~  \text{for}~k=0,1,2,\dots
\end{split}
\end{equation*}

Now, the PDF of $C_m$ can be found by averaging over the PMF of subcarrier collisions as follows:
\begin{equation}\label{eq:f_C_mm2}
\begin{split}
f_{C_m}(x)=& \sum_{\mathbf{k}_m}f_{C_m, \mathbf{K}_m}(x, \mathbf{k}_m) \\ =& \sum_{\mathbf{k}_m}f_{C_m| \mathbf{K}_m}(x | \mathbf{k}_m) p(\mathbf{k}_m).
\end{split}
\end{equation}

Plugging~\eqref{eq:m-hypg} and~\eqref{eq:f_C_mm} into~\eqref{eq:f_C_mm2} results the PDF, and is given in~\eqref{eq:f_{C_m}(x)_final}.
\begin{figure*}
\begin{equation}\label{eq:f_{C_m}(x)_final}
\begin{split}
f_{C_m}(x) =& \sum_{k_{1m}} \sum_{k_{2m}} \cdots \sum_{k_{Nm}} \sum_{k_{fm}} \left\{ \binom{F_f}{k_{fm}}
\binom{F}{F_m^S}^{-1} \prod_{n=1}^{N} \binom{F_n^P}{k_{nm}}
\left ( \frac{\beta_{\min}}{\beta^{NI}} \right )^{\alpha^{NI}k_{fm}}\prod_{n=1}^{N}\left ( \frac{\beta_{\min}}{\beta_n^{I}} \right )^{\alpha_n^{I}k_{nm}} \right.
\\
& \times \left. \sum_{k=0}^{\infty} \frac{\delta_k x^{\sum_{n=1}^{N}\alpha_n^Ik_{nm} + \alpha^{NI}k_{fm}+k-1} \exp\left ( -\frac{x}{\beta_{\min}} \right )}{\beta_{\min}^{\sum_{n=1}^{N}\alpha_n^Ik_{nm} + \alpha^{NI}k_{fm}+k} ~\Gamma\left ( \sum_{n=1}^{N}\alpha_n^Ik_{nm} + \alpha^{NI}k_{fm} +k  \right )}U(x) \right \}. 
\end{split}
\end{equation}
\vspace*{0.3cm}
\\
\rule[0.2cm]{1\textwidth}{0.017cm} 
\end{figure*}

The outage probability is a common performance metric in fading environments. Hence, here we consider
the outage probability of SU capacity in terms of the following measure:
\begin{equation*}
\begin{split}
P_{C_m}^{\mathrm{out}}(\varphi_{\mathrm{th}} )= & Pr\left ( C_m < \varphi_{\mathrm{th}} \right ) \\ = & \int\limits^{\varphi_{\mathrm{th}}}_{0}f_{C_m}(x )  \mathrm{d}x,
\end{split}
\end{equation*}
which is the CDF of the SU capacity over the outage threshold $\varphi_{\mathrm{th}}$ [dB]. Using~\eqref{eq:mos_cdf2} and~\eqref{eq:f_{C_m}(x)_final}, the CDF of $C_m$ can be expressed as
\begin{equation}\label{eq:F_{C_m}(x)_final}
\begin{split}
F_{C_m}(x) & = \sum_{k_{1m}} \sum_{k_{2m}} \cdots \sum_{k_{Nm}} \sum_{k_{fm}} \left\{ \binom{F_f}{k_{fm}}
\binom{F}{F_m^S}^{-1} 
\right. \\
& \quad \left. \times \hspace*{-0.5mm} \prod_{n=1}^{N}\hspace*{-0.5mm} \binom{F_n^P}{k_{nm}}  \left ( \frac{\beta_{\min}}{\beta^{NI}} \right )^{\alpha^{NI}k_{fm}}\prod_{n=1}^{N}\hspace*{-0.5mm} \left ( \frac{\beta_{\min}}{\beta_n^{I}} \right )^{\alpha_n^{I}k_{nm}}
\right. \\
& \quad \times  \hspace*{-0.5mm}\left. \sum_{k=0}^{\infty} \delta_k \mathcal{P} \hspace*{-0.5mm}\left ( \sum_{n=1}^{N}\alpha_n^Ik_{nm} + \alpha^{NI}k_{fm}+k, \frac{x}{\beta_{\min}} \right )\hspace*{-1mm}\right \}\hspace*{-0.5mm}.
\end{split}
\end{equation}

\section{Asymptotic Analysis of Multiuser Diversity}\label{sec:multiuser}

In this section, the gain of multiuser diversity by employing opportunistic scheduling  is investigated. In conventional systems, the multiuser diversity gain is attributed to {\it channel gains} only. However, in the  proposed scheme, we additionally benefit from the randomness of {\it the number of subcarrier collisions}.  Assuming all $M$ SUs are accessing the $F$  available subcarriers to randomly allocate their subcarriers,\footnote{It is assumed that  no collisions occur among  the subcarriers of SUs.} the SU, which provides the best instantaneous capacity, is selected as:
\begin{equation*}
C_{\max}  = \max_{m \in [1, M]} C_m.
\end{equation*}

For fairness in the selection phase of the best SU, assume that each SU's data rate is the same, i.e., each SU requests for the same number of subcarriers, $F_m^S=F^S,~m=1,\dots,M$.  Then, by using order statistics,  the PDF of $C_{\max}$  is expressed as
\begin{equation}\label{eq:f_{C_{max}}(x)}
f_{C_{\max}}(x)= Mf_{C_m}(x)F_{C_m}(x)^{M-1}.
\end{equation}

Plugging~\eqref{eq:f_{C_m}(x)_final} and~\eqref{eq:F_{C_m}(x)_final} into~\eqref{eq:f_{C_{max}}(x)}, the PDF of $C_{\max}$ can be obtained. Nonetheless, using $\int_{-\infty}^{\infty}x f_{C_{\max}}(x)\mathrm{d}x$ is intractable to find the mean of  $C_{\max}$. Even if we can carry out such a calculation, it will hardly provide any insights to fully understand the impacts of the main parameters on the capacity using the resulted expression. Therefore,  we asymptotically analyze the capacity to understand the effects of system  parameters and multiuser diversity gain in CR systems with spectrum sharing feature.

\begin{theorem}\label{teo:max}
As the number of SUs $M$ goes to infinity, the average capacity of $C_{\max}$  converges to
\begin{equation*}
\mathbb{E}\left [C_{\max}\right] = b_M +  E_1 a_M,
\end{equation*}
where $E_1=0.5772\dots$ is Euler's constant~\cite{david1981order}, and $a_M =\left [M f_{C_m}(b_M)\right ]^{-1}$ and $b_M =F_{C_m}^{-1}(1- 1/M)$.

 Without loss of generality assuming a {\it single} PU case, i.e., $n\in[1,N]$, then $b_M$ is given by
\begin{equation*}
\begin{split}
b_M & =   F_{C_m^1}^{-1}\left (1- \frac{1}{M}\right )\\ & =  \mathcal{Q} \sum\limits_{k=0}^{\infty}\delta_k  \mathcal{P}^{-1}\left ( \Delta+k,\frac{1- \frac{1}{M}}{\hat{\beta}_{\min}} \right ),
\end{split}
\end{equation*}
where $\mathcal{P}^{-1}(\cdot, \cdot)$ stands for the inverse regularized  incomplete Gamma function. Unfortunately, there is no closed form expression for this special function. Therefore, it can be evaluated numerically by using build-in functions in some well-known computational softwares such as  MATLAB$\textsuperscript{\textregistered}$ and MATHEMATICA$\textsuperscript{\textregistered}$.\footnote{It is worth to note that by using~\cite[6.5.12 \& 13.5.5]{abramowitz1964handbook}, the  regularized  incomplete Gamma function can be approximated as $\mathcal{P}\left ( u,v \right )= \frac{v^u}{u\Gamma (v)}~ _{1}F_1\left ( u; 1+u;-v \right )  = \frac{v^u}{u\Gamma (u)}$ as $v\to 0$, where $ _{1}F_1\left ( \cdot; \cdot ;\cdot \right )$ is confluent hypergeometric function~\cite{Dohler2006inverse}. Hence, its inverse can be obtained.} Additionally, $\hat{\beta}_{\min} = \min \{ \beta^{I}_n, \beta^{NI}\}$, $\Delta= \alpha^I_n k_{nm} + \alpha^{NI} k_{fm}$ and $\mathcal{Q}$ takes the form:
\begin{equation*}
\begin{split}
\mathcal{Q}  = & 
\binom{F}{F_m^S}^{-1}  \sum_{k_{nm}=0}^{F_m^S}  \binom{F_n^P}{k_{nm}}
\binom{F - F_n^P}{k_{fm}}
\\
&  \times \left (\frac{\hat{\beta}_{\min}^{\Delta}} {(\beta^I_n)^{\alpha^I_n k_{nm} } (\beta^{NI})^{\alpha^{NI} k_{fm} }} \right ),
\end{split}
\end{equation*}
where in  considering a practical scenario, it is assumed that $ F_m^S + F_n^P \le F $ and $ F_m^S \le F_n^P$. Hence, the support region for the number of subcarrier collisions is $k_{nm}=0,1,\dots,{F_m^S}$.
\end{theorem}

\begin{IEEEproof}
We start with the following {\it Lemma}.
\begin{lemma}[Distribution of Extremes~\cite{ david1981order}]\label{lem:extreme}
Let $z_1,\dots, z_M$ be  i.i.d. RVs with
absolutely continuous common CDF, $F(z)$, and PDF, $f(z)$, satisfying these conditions: $F(z)$ is less than 1
for all $z$,   $f(z)>0$  and is differentiable. If  the growth function
$g(z)= (1-F(z))/f(z)$    satisfies the  von Mises' sufficient condition:
\begin{equation}\label{eq:lim}
\lim_{z\to \infty } g(z) =c >0,
\end{equation}
then $F(z)$ belongs to the domain of attraction of the Gumbel distribution. In other words, $[\max_{1\le k \le M} z_m - b_M] /a_M$
converges in distribution to the Gumbel-type limiting distribution:
\begin{equation*}
G(x)=\exp \left (-e^{-z} \right)\quad  , \; \;  -\infty<z<\infty ~.
\end{equation*}

  Thus, the maximum of $M$ such i.i.d. RVs grows like $b_M$, also termed   position parameter.
The parameter $b_M$ is given by $b_M =F^{-1}(1- 1/M)$, and the scaling factor $a_M$ is given by $a_M=g(b_M)  =\left [M f(b_M)\right ]^{-1}$.
\end{lemma}

The PDF and CDF of $C_m$  for a {\it single} PU are given, respectively, by
\begin{equation}\label{eq:f_C_m_1}
f_{C_m^1}(x) = \mathcal{Q} \sum_{k=0}^{\infty} \frac{\delta_k x^{{\Delta+k-1} } e^{ -{x/\hat{\beta}_{\min}}}}{\hat{\beta}_{\min}^{{\Delta+k}}~\Gamma\left ( \Delta +k \right )}U(x),
\end{equation}
\begin{equation}\label{eq:F_C_m_1}
F_{C_m^1}(x) = \mathcal{Q}
\sum_{k=0}^{\infty}\delta_k \mathcal{P}\left ( \Delta +k, \frac{x}{\hat{\beta}_{\min} } \right ),
\end{equation}
with the coefficients calculated iteratively as
\begin{equation*}
\begin{split}
&  \delta_0 =1 \\
& \delta_k = \frac{1}{k+1} \sum\limits_{i=1}^{k+1}\left [ \alpha^I_n k_{nm} \left ( 1- \frac{\hat{\beta}_{\min}}{\beta^I_n} \right )^i 
\right.
\\& \left. ~~~~~ +\alpha^{NI}  k_{fm} \left ( 1- \frac{\hat{\beta}_{\min}}{\beta^{NI}} \right )^i \right ]\delta_{k+1-i}   ~~ \text{for}~k=0,1,2,\dots
\end{split}
\end{equation*}

From {\it Lemma}~\ref{lem:extreme}, plugging~\eqref{eq:f_C_m_1} and~\eqref{eq:F_C_m_1} into~\eqref{eq:lim} yields
\begin{equation}\label{eq:limssss}
 \lim_{x \to \infty}\frac{1- F_{C_m^1}(x) }{f_{C_m^1}(x)}
=\hat{\beta}_{\min}>0.
\end{equation}

The respective intermediate steps in the evaluation of~\eqref{eq:limssss} are depicted in Appendix~\ref{app:lims}.
Hence, it  belongs to an attraction domain of Gumbel-type with limiting CDF:
\begin{equation*}
\hat{F}_{C_{\max}}(x) = \exp\left ( \hspace*{-1mm}-\exp\left (-\frac{x-b_M}{a_M}  \right ) \right ).
\end{equation*}

Then, the limiting PDF of $C_{\max}$ is
 \begin{equation*}
\hat{f}_{C_{\max}}(x) = \hspace*{-1mm}\frac{1}{a_M}\exp\left (\hspace*{-1mm}-\frac{x-b_M}{a_M}  \right )\exp\left (\hspace*{-1mm} -\exp\left (\hspace*{-1mm} -\frac{x-b_M}{a_M}  \right )\hspace*{-1mm} \right )\hspace*{-0.5mm}.
\end{equation*}

Therefore, using $\mathbb{E}\left [C_{\max}\right] = \int_{-\infty}^{\infty}x \hat{f}_{C_{\max}}(x) \mathrm{d}x$, the desired result can be readily obtained.
\end{IEEEproof}

In the proof stage, it came to our attention that, to the best of the authors' knowledge, there is no result reported in the literature for the limiting distribution of RVs that follows Moschopoulos PDF. Therefore, the following novel result can be stated.

\begin{corollary}
Let $\left \{ X_r \right \}_{r=1}^\mathcal{R}$ be the set of $\mathcal{R}$ i.i.d. RVs that follow Moschopoulos PDF and CDF~\cite{moschopoulos1985distribution}, and  $Y = \max \left \{ X_1, X_2, \dots, X_{\mathcal{R}} \right \}$, then the  limiting distribution of the CDF  of $Y$ belongs to the domain of attraction of Gumbel distribution as $\mathcal{R}$ converges to infinity.
\end{corollary}
\begin{IEEEproof}
It is immediate to see this result from the results presented in the proof of {\it Theorem}~\ref{teo:max}.
\end{IEEEproof}

The results obtained so far will help us to asymptotically analyze the  scheduling of SUs' subcarriers  in the following section.

\section{Centralized Sequential and Random Subcarrier Allocation}\label{sec:scheduling}

\subsection{Sum Capacity of SUs with Multiuser Diversity}

In this section, a cognitive communication set-up involving multiple SUs that assume a  random allocation method is studied. Recall that  due to random allocation scheme, there can be the collisions among  the subcarriers of SUs in addition to those that are used by PUs. These collisions will decrease the system performance severely. To overcome this challenge, we propose an efficient algorithm
that {\it sequentially} and {\it randomly} allocates  SUs' subcarrier sets in a centralized manner by maintaining the orthogonality among the allocated subcarrier sets. Such an assignment can be thought of as the downlink scenario where the SBS performs the random assignment of subcarriers.
Furthermore, to benefit from the multiuser diversity gain, the opportunistic scheduling method is employed in the algorithm [See Table~\ref{alg:seq}], where it is assumed  only a single PU. The multiple PUs case is a  straightforward extension.  In the selection step of the best SU,  to preserve the fairness among the users, it is assumed that the data rate requirements of all SUs are the same, i.e., the individual numbers of subcarrier requirements are equal.


\begin{table*}
\small
  \caption{ALGORITHM: CENTRALIZED SEQUENTIAL AND RANDOM SUBCARRIER ALLOCATION}
\begin{center}
  \begin{tabular}{l}
{\fbox{\parbox{6.35in}{
\begin{enumerate}
  \item {\em Initialization}
    \begin{itemize}
      \item Assume $F_m^S = F^S~\forall m \in [1,M]$ and a single PU is available, $n=1$.
      \item Set the number of available subcarriers to $F$ and index $t=1$.
    \end{itemize}
  \item {\em Subcarrier assignment step}
    \begin{itemize}
      \item Randomly sample a set of subcarriers, $F_t^R$, with cardinality of $F^S$ from set $F$:  $k_{nm}\sim \texttt{HYPG}(F^S,~ F_n^P, ~ F)$.
      \item  Assign the set $F_t^R$  to all $M-t+1$ SUs.
    \end{itemize}

  \item {\em Capacity calculation  step}
    \begin{itemize}
    \item For $m=1,\dots,M-t+1$,  SUs evaluate their capacities with the given random set of subcarriers:~$C_{m} \big | F_t^R$.
    \item SUs send {\it feedback} for the calculated capacities to the {\it central control entity} (SBS or CR Network Manager).

    \end{itemize}
  \item {\em Selection step}
    \begin{itemize}
      \item  Choose the SU that provides the best capacity:
\\ If $t=1$ then $m_t^*= \underset{{m\in[1,M]}}{\arg\max}\left (C_{m} \big | F_t^R  \right )$\\
else
$m_t^*= \underset{{m\in\left[1,M\right]\backslash \left[m_1^*, m_{t-1}^*\right]}}{\arg\max}  \left (C_{m} \big | F_t^R  \right ) \quad \text{for}~ t=2,\dots, \hat{M}$.
    \end{itemize}

  \item {\em Updating the subcarrier sets step}
    \begin{itemize}
      \item  Remove the sampled (total of collided and collision-free) subcarriers from the available set of subcarriers:\\
 $F \leftarrow F- F_t^R$.

\item  Set $t\leftarrow t+1$ and go to Step 2 until $t= \hat{M}$.
    \end{itemize}

  \item {\em Sum capacity evaluation step}
    \begin{itemize}
        \item Compute sum capacity of SUs: $C_{\mathrm{sum}}=\sum\limits_{t=1}^{\hat{M}} C_{m_t^*}$.
    \end{itemize}

\end{enumerate}
}}}
  \end{tabular}
\end{center}
\label{alg:seq}
\end{table*}


The algorithm can be summarized as follows. A randomly chosen set of subcarriers $F_t^R$ from the set $F$ is assigned to the available SUs. The first SU, which provides the best capacity, is selected among $M$ SUs, then the selected subcarriers $F_t^R$  (total of collided and collision-free subcarriers) are removed from the set $F$. In the next stage, another randomly chosen set of subcarriers $F_t^S$ from the {\it updated} set $F$ is allocated to the rest of SUs. The second best SU is selected among the $M-1$ SUs, and similarly the subcarrier set $F$ is updated by removing the new set $F_t^R$. This sequential selection continues until it reaches the total number of the best $\hat{M}$  SUs, with $\hat{M}\le M$. It is evident to observe that the multiuser diversity is attributed only to the randomness of the channel gains. Furthermore, some of the essential points in the algorithm can be highlighted as follows:
\begin{itemize}
\item In {\it step} 2: The PMF of the number of subcarrier collisions follows a {\it hypergeometric distribution} due to the random selection of subcarriers set $F_t^R$ from the available set of subcarriers $F$.

\item In {\it step} 4: The selection of the best SU is performed based on the capacity feedbacks obtained from the SUs.

\item In {\it step} 5: Removing the randomly sampled subcarriers $F_t^R$ from the available set of subcarriers $F$ means that both collided and collision-free subcarriers are subtracted from set $F$ (since $F_t^R = k_{nm}+k_{fm}$), i.e., $F \leftarrow F- F_t^R  \Leftrightarrow  F \leftarrow F- k_{nm} - k_{fm}$ and $F_n^P \leftarrow F_n^P- k_{nm}$.
In other words, since $F = F_n^P + F_f$, where $F_f$ stands for the number of free subcarriers, the subcarriers that are occupied by the PU $F_n^P$ in the set $F$  are automatically updated when the randomly sampled set of subcarriers $F_t^R$ is removed from the  set $F$. Hence, the orthogonality among the subcarriers of SUs is maintained.

\end{itemize}

\begin{theorem}
The sum capacity of $\hat{M}$ selected SUs in the centralized sequential and random scheduling algorithm for $M \gg  \hat{M}$ is approximated\footnote{Since $\mathbb{E}\left [ C_{m_1^*}\right]\ge\mathbb{E}\left [ C_{m_j^*}\right], \forall j\in\left [1,\hat{M}\right]$, it can also be considered as a  tight upper bound  for $M \gg  \hat{M}$ as  $\mathbb{E}\left [C_{\mathrm{sum}}\right]\le\hat{M} \mathbb{E}\left[ C_{m_1^*}\right]$.} by
\begin{equation*}
\mathbb{E}\left [C_{\mathrm{sum}}\right]\approx\hat{M} \mathbb{E}\left [ C_{m_1^*}\right],
\end{equation*}
and as $M\to \infty$,  it converges to
\begin{equation*}
\mathbb{E}\left [C_{\mathrm{sum}}\right]= \hat{M} \left [ b'_{M} +  E_1 a'_{M} \right],
\end{equation*}
where $m_1^*$ is the index of the first selected best SU and defined as $m_1^*= \underset{{m\in[1,M]}}{\arg\max}~C_m$. Further, $a'_{M}$ and $b'_{M}$ can be readily obtained by following the same approach as in {\it Theorem}~\ref{teo:max} considering the fact that the multiuser diversity is only ascribed to channel randomness not the random subcarrier assignment. It is noteworthy to state that the sum capacity scales linearly with the number of selected SUs.
\end{theorem}

\begin{IEEEproof}
The scheduler selects the SUs according to the following rule:
\begin{equation*}
m_j^*= \underset{{m\in[1,M]\backslash \left [m_1^*, m_{j-1}^*\right]}}{\arg\max}~C_m \quad \text{for}~ j=2,\dots, \hat{M}~,
\end{equation*}
which means that the selected SU(s) are ignored in the selection step of remaining users.
Then, the sum capacity of  selected SUs  is defined as
\begin{equation}\label{eq:algorithm1}
C_{\mathrm{sum}}=\sum_{j=1}^{\hat{M}} C_{m_j^*}.
\end{equation}

For large $M$ such that $M \gg  \hat{M}$,  it is immediate to observe that
\begin{equation}\label{eq:algorithm2}
\mathbb{E}\left [C_{m_j^*}\right]\approx \mathbb{E}\left [C_{m_1^*}\right]\quad \forall j\in\left [1,\hat{M}\right].
\end{equation}

This approximation is valid since removing the selected SUs does not considerably impact the mean of the rest of the selected SUs for $M \gg  \hat{M}$, i.e., the maxima of $M$ RVs and $M-\hat{M}$ RVs are approximately the same  for $M \gg  \hat{M}$, so their averages are  approximately the same.
Hence, plugging~\eqref{eq:algorithm2} into~\eqref{eq:algorithm1} yields the desired result.
\end{IEEEproof}


\subsection{Sum Capacity of SUs without Opportunistic Scheduling}\label{}

In order to investigate the performance of our proposed algorithm due to multiuser diversity gain, the performance of the centralized sequential subcarrier scheduling without employing the opportunistic scheduling method is analyzed  in this section, i.e.,  the multiuser diversity of SUs is not maintained. Therefore, the sum capacity of any {\it arbitrarily}  $\hat{M}$ selected SUs (among $M$ SUs) can be expressed  as
\begin{equation*}
C_{\mathrm{sum}}^\mathrm{a} = \sum_{m=1}^{\hat{M}} C_m.
\end{equation*}

Recalling the upper and lower bounds on the average capacity of a single SU $C_m$, one can conclude that the average sum rate of the SUs  scales linearly with the number of selected SUs ($\hat{M}$). Mathematically speaking, $\mathbb{E}\left [C_{\mathrm{sum}}^\mathrm{a}\right] = \hat{M} \mathbb{E}\left [C_m\right]$.

During the sequential scheduling of SUs' subcarriers, the PMF of the number of subcarrier collisions can be obtained as a special case of the following result.

\begin{figure}[t]
    \begin{center}
     \includegraphics[width=0.47\textwidth]{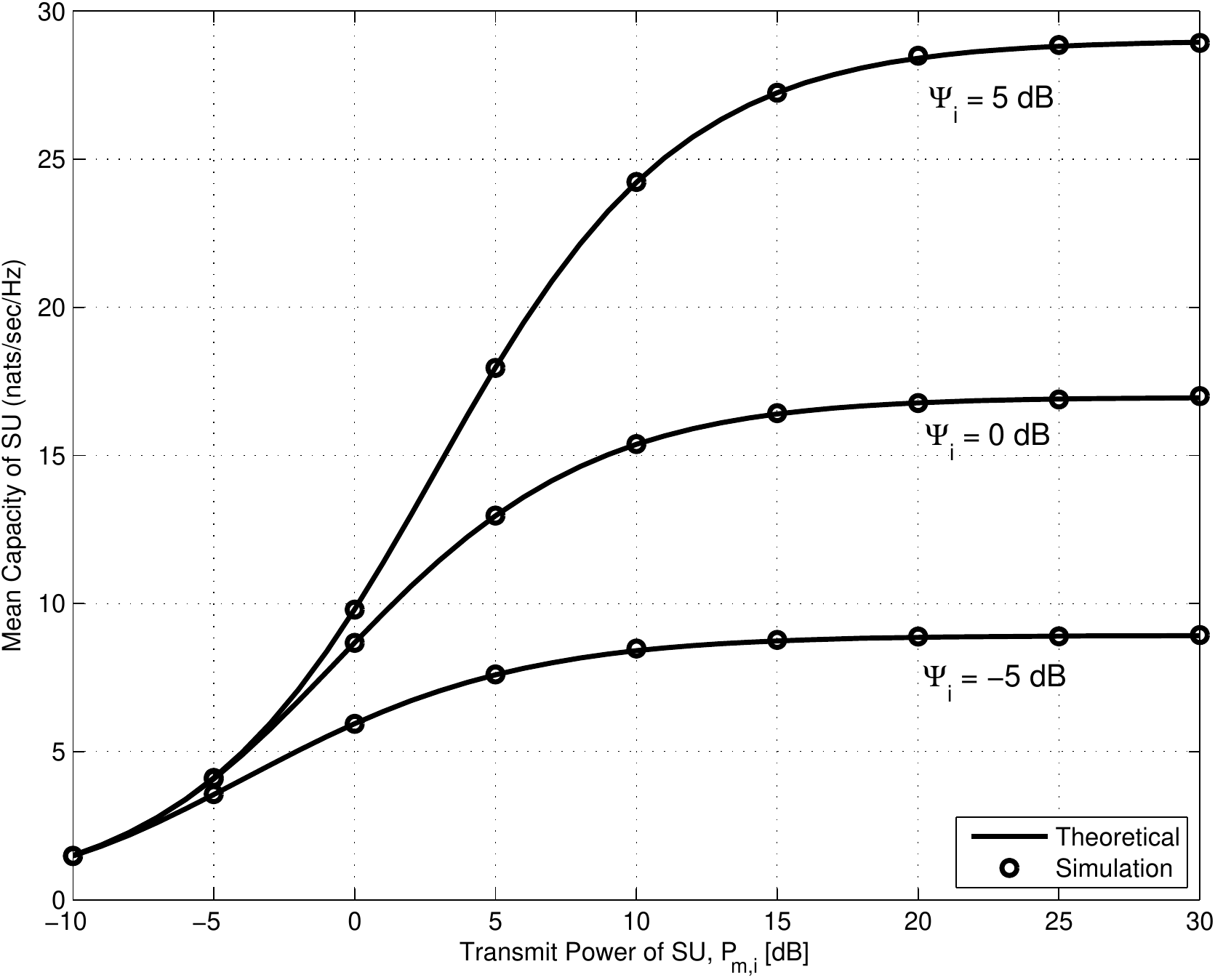}
 \caption{SU mean capacity versus the transmit power $P_{m,i}$ with different IT $\Psi_i$ values for $F_m^S= 20$, $F_n^P= 30$, $F= 128$ and $P_{n,i}=10$ dB.}
 \label{fig:fig4}
    \end{center}
\end{figure}

\begin{proposition}\label{teo:PMF}
The~PMF of the number of subcarrier collisions for the $m$th SU in the presence of $N$ PUs, when assigning the subcarriers sequentially to preserve the orthogonality between SUs' subcarriers, is given by
\begin{equation*}
p(\mathbf{k}_{m})=\sum_{\mathbf{k}_{1}}\sum_{\mathbf{k}_{2}}\cdots\sum_{\mathbf{k}_{m-1}}p(\mathbf{k}_{1}, \mathbf{k}_{2}, \dots, \mathbf{k}_{m}),
\end{equation*}
where the joint PMF is
\begin{equation*}
\begin{split}
& p \left (\mathbf{k}_1, \mathbf{k}_2, \dots, \mathbf{k}_{m} \right )
= \\
&  \left [ \binom{F_f}{k_{f1}}
 \binom{F}{F_1^S}^{-1} \prod_{n=1}^{N}\binom{F_n^P}{k_{n1}} \right ]
\prod_{r=2}^{m}\left \{ \binom{F_f -  \sum_{j=1}^{r-1}k_{fj}}{k_{fr}}
 \right.
\\
& \times \left. \binom{F -\mathbf{1}^\mathrm{T}\left(\sum_{j=1}^{r-1}\mathbf{k}_j\right )}
{F_r^S}^{\hspace*{-1mm}-1}  
\prod_{n=1}^{N}\binom{F_n^P - \sum_{j=1}^{r-1}k_{nj}}{k_{nr}} \right \}.
\end{split}
\end{equation*}

The mean and support  of $k_{nm}$ are given, respectively, by
\begin{equation*}
\mathbb{E}\left [k_{nm}\right] = \frac{F_m^S \left ( F_n^P - \sum\limits_{j=1}^{m-1} \mathbb{E}\left [k_{nj}\right]\right )}{F - \sum\limits_{j=1}^{m-1} F_j^S },
\end{equation*} 
\begin{equation*}
\begin{split}
&  \left \{\mathbf{k}_m :\sum_{n=1}^{N}k_{nm} + k_{fm} =F_m^S  ~\text{and}~  k_{nm}\in  \Bigg (  F_m^S +  F_n^P   \right. \\
& ~~ \left.  - \sum_{j=1}^{m-1}k_{nj} -F \Bigg )^+ , \dots, \min \left \{   F_j^S, F_n^P - \sum_{j=1}^{m-1}k_{nj} \right \}  \right \}.
\end{split}
\end{equation*} 
\end{proposition}
\begin{IEEEproof}
The proof is given in Appendix~\ref{sec:PMF}.
\end{IEEEproof}

\begin{figure}[t]
    \begin{center}
   \includegraphics[width=0.47\textwidth]{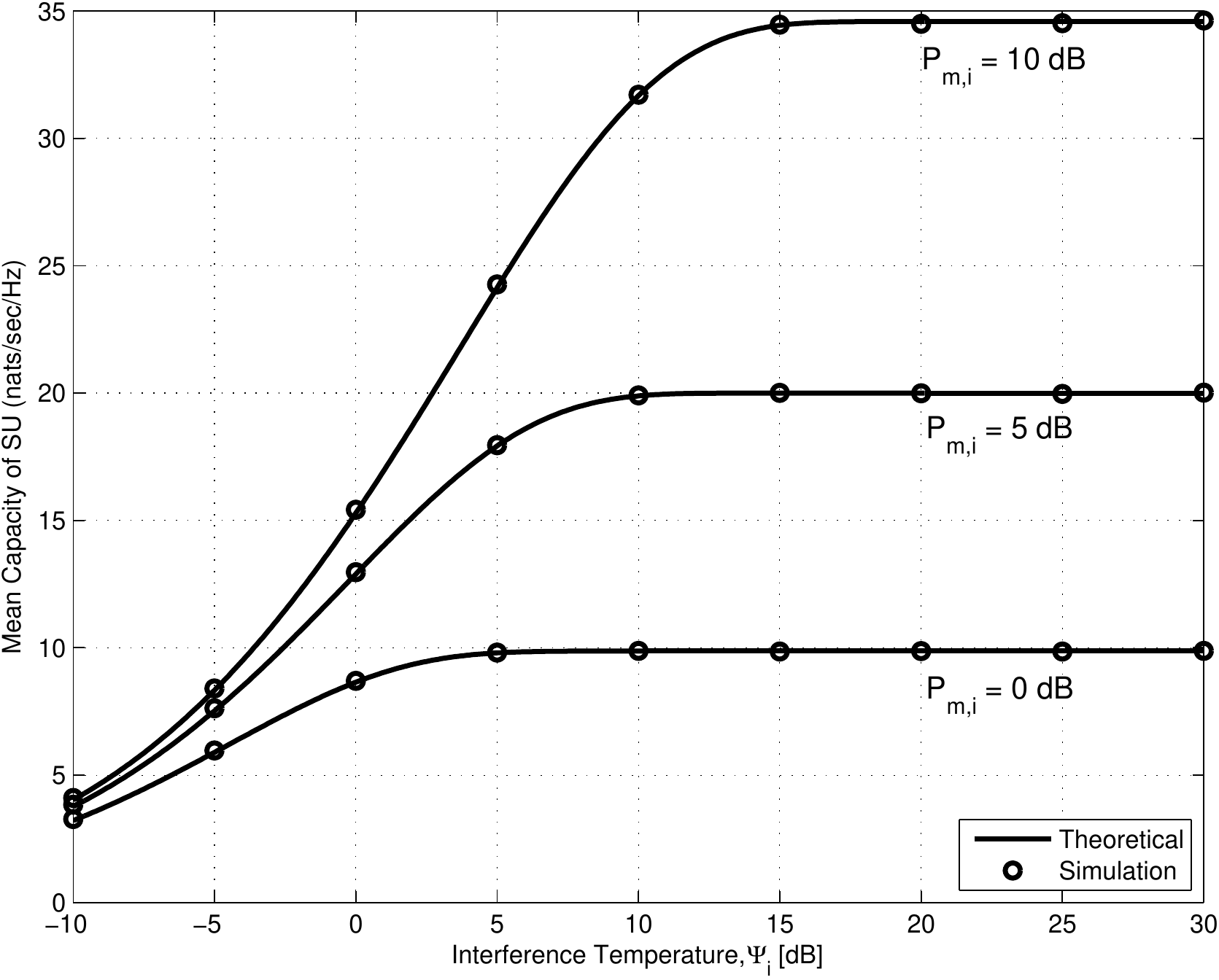}
 \caption{SU mean capacity versus the interference temperature $\Psi_{i}$ with different transmit power $P_{m,i}$ values for $F_m^S= 20$, $F_n^P= 30$, $F= 128$ and $P_{n,i}=10$ dB.}
 \label{fig:fig5}
    \end{center}
\end{figure}

\section{Numerical Results and Simulations}\label{sec:num_res}

In this section, numerical and simulation results are presented  to confirm the analytical results and investigate the impact of various system parameters in CR spectrum sharing networks.  First, the effect of peak transmit power of SU $P_{m,i}$ (in dB) on the
average capacity (in nats per second per hertz) is shown for different values of IT  values $\Psi_i$ in Figure~\ref{fig:fig4}. Unlike the conventional systems, the SU average capacity is here saturated after a certain
value of peak SU transmit  power because of the IT constraint in  spectrum sharing systems.
In Figure~\ref{fig:fig5}, the SU mean capacity against the IT constraint is presented. It turns out that the analytical results agree well with the simulation results. The results shown in Figures~\ref{fig:fig4} and~\ref{fig:fig5} are in the presence of a single PU, i.e., $n\in [1,N]$, and the number of subcarriers in sets $F$, $F_m^S$ and $F_n^S$ are chosen arbitrarily.\footnote{The unit AWGN noise variance  is used ($\eta = 1$) in all the following figures.}  A common observation for both   Figures~\ref{fig:fig4} and~\ref{fig:fig5}  is that the saturation level of
capacity increases as the IT constraint relaxes, and the capacity keeps growing until a saturation point as the transmit power of SU increases as expected.
It can also be underlined from Figure~\ref{fig:fig4} that the capacity gain due to relaxation in the IT constraint disappears at low SU transmit power. Therefore, in the high transmit power or SINR regime, the impact of IT relaxation differs significantly. Similarly, the same effect can be observed for the results in Figure~\ref{fig:fig5}.

\begin{figure}[t]
    \begin{center}
    \includegraphics[width=0.47\textwidth]{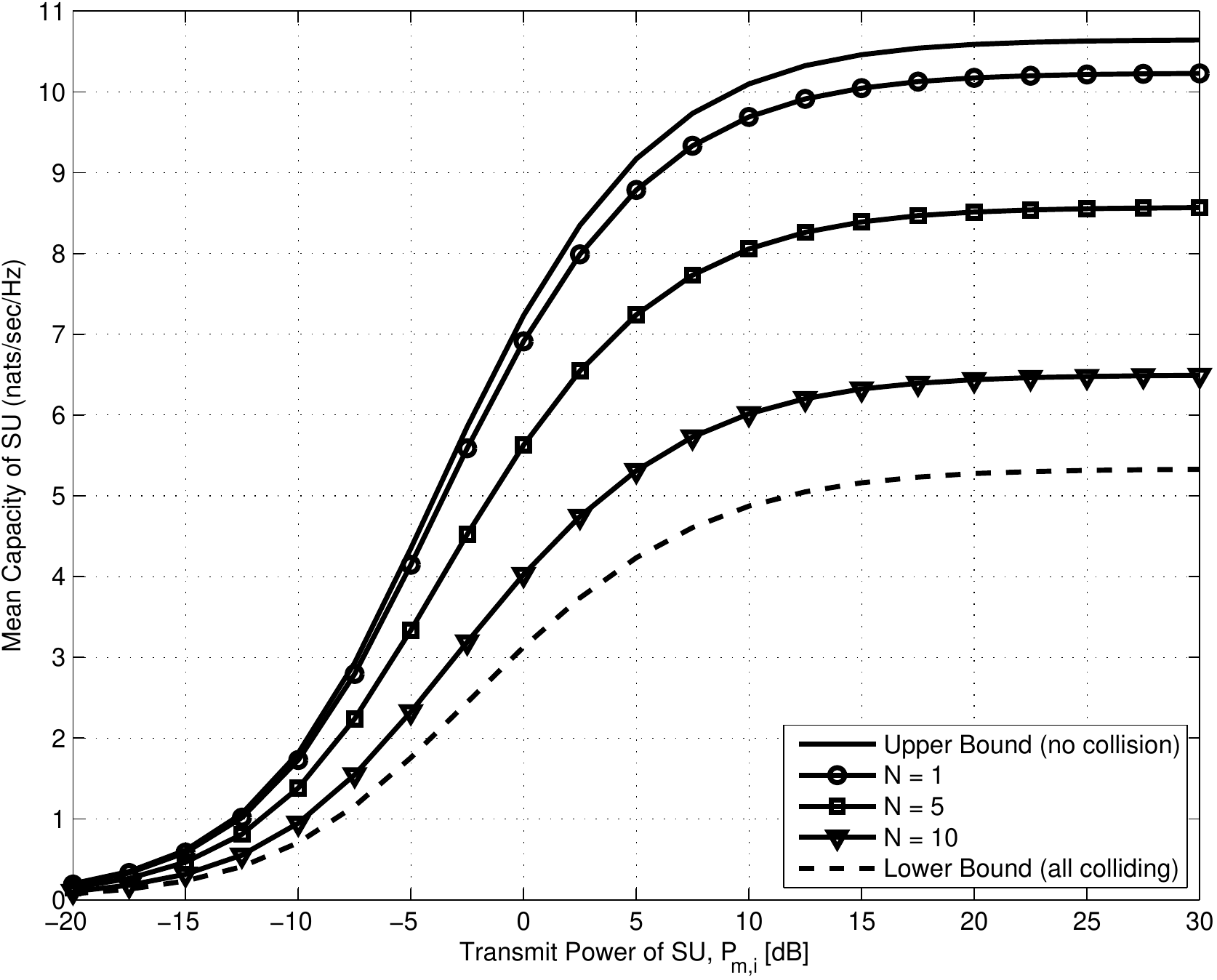}
 \caption{SU mean capacity versus the transmit power $P_{m,i}$ with different number of PUs $N$ for $F_m^S= 20$,  $F= 128$, $F_n^P= 10$, $P_{n,i}=5$ dB for $n=1,\dots,N$ and $\Psi_i = -5$ dB.}
 \label{fig:fig6}
    \end{center}
\end{figure}

Consider now the practical scenario when there are multiple PUs available. Therefore, the number of free subcarriers in the available set $F$ is smaller than that of the single PU case. The  SU mean capacity against  peak transmit power $P_{m,i}$ in the presence of multiple PUs is shown in Figure~\ref{fig:fig6}. In order to illustrate the effects of multiple PUs, during the simulations, it is assumed that the number of subcarriers and the transmit power of all PUs are the same, $P_{n,i}=5$dB and $F_n^P=10$ for $n=1,\dots,N$, respectively.
Since the number of subcarrier collisions in the presence of multiple PUs follows a multivariate hypergeometric distribution, the multivariate hypergeometric random variates are generated by using the sequential  method given in~\cite[p.~206]{gentle2003random}. It can be observed that increasing the number of PUs degrades the performance of SU as expected. In addition, as the number of PUs decreases, i.e., the number of unoccupied subcarriers  increases, the average capacity of SUs converges to the upper bound, where all SU's subcarriers are collision-free.  On the other hand, the lower bound of the average capacity indicates that all SU's subcarriers are colliding.

\begin{figure}[t]
    \begin{center}
    \includegraphics[width=0.47\textwidth]{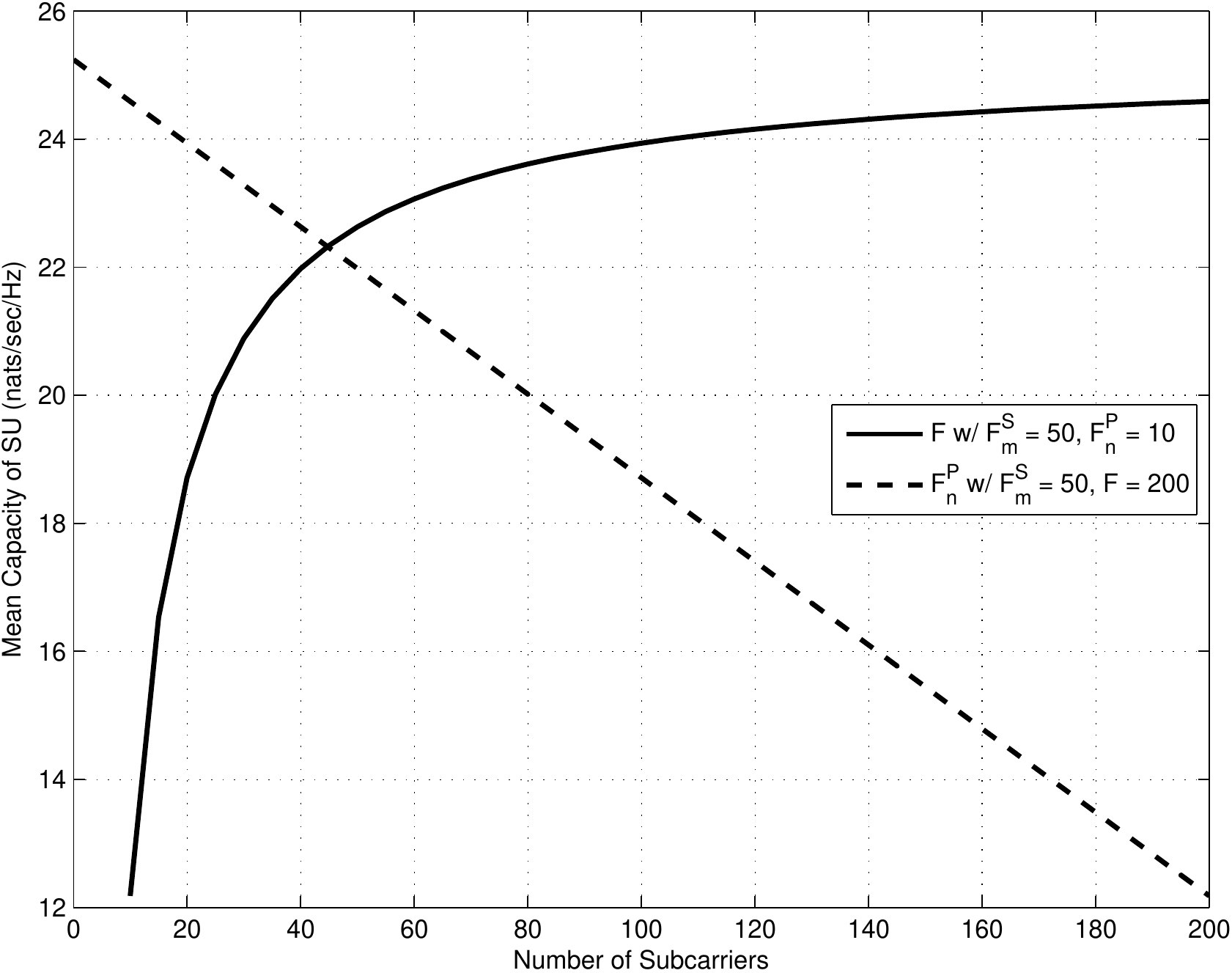}
 \caption{SU mean capacity versus the number of subcarriers $F$ and $F_n^P$, for  $P_{n,i}= 5$ dB,  $P_{m,i}=10$ dB and $\Psi_i = -5$ dB.}
 \label{fig:fig7}
    \end{center}
\end{figure}

Figure~\ref{fig:fig7} shows how the SU average capacity scales with the number of subcarriers in sets $F$ and $F_n^P$, respectively, where the single PU case is assumed. As the number of available subcarriers increases for a fixed number of SU's and PU's subcarriers, the SU mean capacity asymptotically converges to the limit point given in~\eqref{eq:rate_conv}, where the rate of convergence is logarithmic. It is immediate to see that the SU average capacity scales as $\Theta  \left( 1 +  1/F \right )$,  $\Theta  \left( F_m^S \right )$ and  $\Theta  \left( 1 - F_n^P \right )$, as proved in {\it Corollary}~\ref{cor:scaling}.

\begin{figure}[t]
    \begin{center}
     \includegraphics[width=0.47\textwidth]{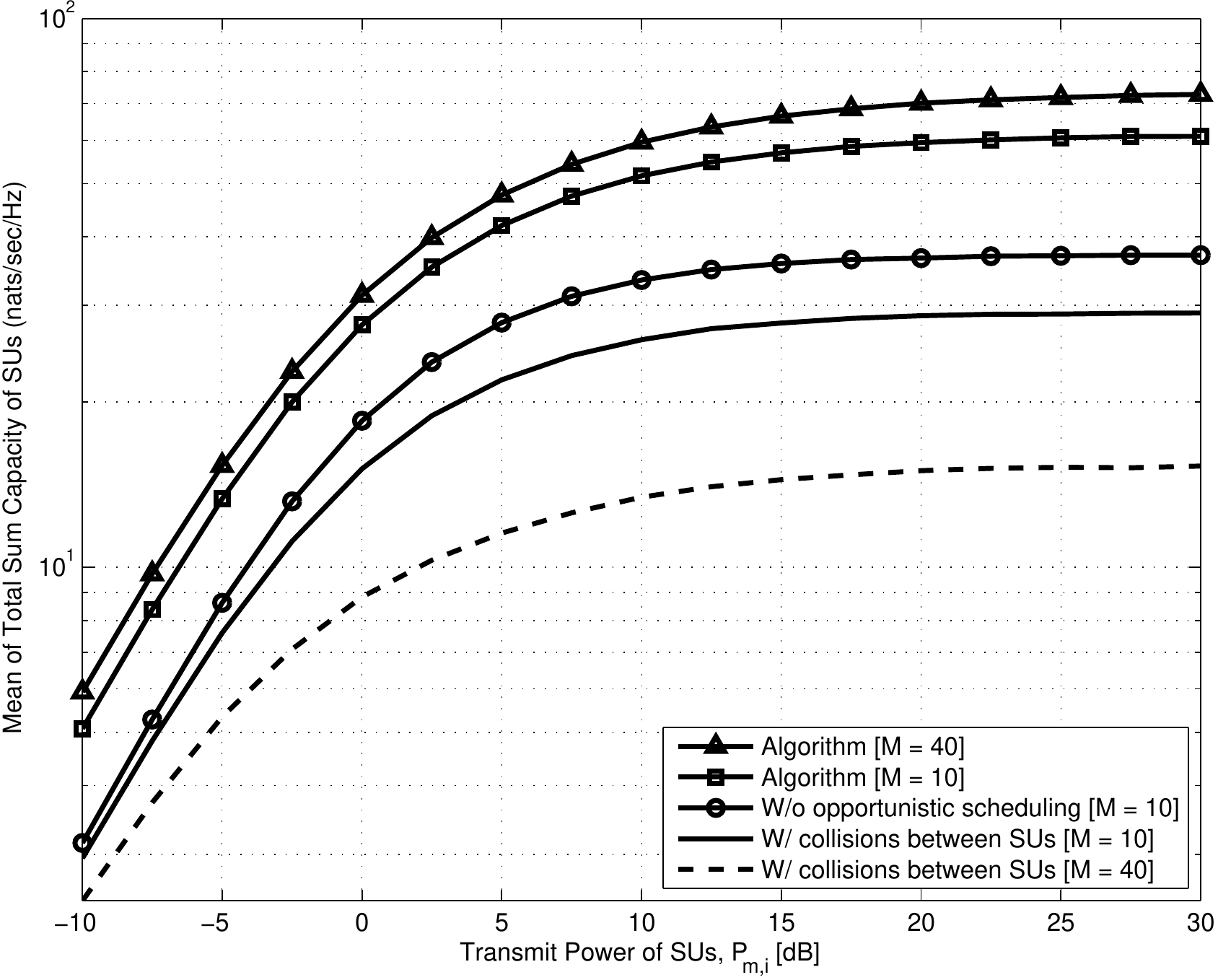}
 \caption{Sum capacity of $\hat{M}=5$ selected SUs versus the transmit powers $P_{m,i}$ for $F_m^S= 10$, $m=1,\dots,M$, $F_n^P= 40$, $F= 100$,  $P_{n,i}=10$ dB and $\Psi_i=0$ dB.}
 \label{fig:fig8}
    \end{center}
\end{figure}

The performance of the proposed centralized algorithm with and without the opportunistic scheduling is simulated and shown in Figure~\ref{fig:fig8}. The results for the algorithm with multiuser diversity are in the presence of $M=10$ and $M=40$ SUs and among them $\hat{M}=5$ SUs are selected using the opportunistic selection method. Also,  $\hat{M}=5$ SUs are selected when no opportunistic scheduling method is employed. Note also that without opportunistic scheduling, the number of SUs $M$ does not affect the sum capacity of $\hat{M}$ selected SUs. Therefore, this scheme is not plotted for different  numbers of SUs ($M$).
One can observe that the effect of multiuser diversity manifests into the fact that an increase in the number of SUs $M$ results in higher capacity in the proposed algorithm.
Furthermore, in order to reveal the impact of collisions between SUs subcarriers on the sum capacity of any arbitrarily $\hat{M}$ selected SUs, we simulate the performance of  $\hat{M}=5$ selected SUs in the presence of $M=10$ and $M=40$ SUs in the secondary network when no centralized algorithm with opportunistic scheduling is  employed. In other words, the orthogonality among  the subcarriers of SUs is not maintained, and the multiuser diversity gain is not exploited. Hence, there can be the collisions between the subcarriers of any SU with the rest of the SUs in the secondary network  in addition to those that are utilized by PU. This scheme could be considered as the worst case scenario, where the collisions among  the SUs' subcarriers  severely affect the performance due to high probability of interference level among SUs as shown in Figure~\ref{fig:fig8}.

\section{Conclusions}\label{sec:conc}

This paper studied the performance of OFDM-based CR systems with spectrum sharing feature using a random subcarrier allocation method. The subcarrier collision models for single and multiple PU(s) are shown to assume univariate and multivariate hypergeometric distributions, respectively.  The expressions of SU average capacity for both general and Rayleigh channel fading models are presented. It turns out that the closed-form expression for the instantaneous SU capacity in the presence of Rayleigh channel fading is intractable. Therefore, the  Gamma approximation of the SU capacity expression is obtained by employing the moment matching method and Moschopoulos PDF representation for a sum of independent but not necessarily Gamma distributed RVs. Through the asymptotic analysis of SU mean capacity, it is found that the capacity scales with the number of subcarriers as $\Theta  \left( 1 +  1/F \right )$,  $\Theta  \left( F_m^S \right )$ and  $\Theta  \left( 1 - F_n^P \right )$. The asymptotic analysis of capacity assuming an opportunistic selection method is investigated by using extreme value theory.
When multiple SUs are randomly allocated the subcarriers, the primary issue that causes drastic performance degradation is the collision(s)  among  their subcarrier sets. In order to prevent such a situation, a  centralized algorithm was  developed to sequentially assign orthogonal subcarrier sets to SUs based on a random allocation scheme while  benefiting from the multiuser diversity gain for maximum SUs sum rate.
Besides, it is found that the extreme value limiting distribution of RVs that follow the Moschopoulos PDF belongs to the domain of attraction of the Gumbel distribution.


\appendices
\section{Proof of Theorem~\ref{teo:avg_cap_single_PU}}\label{proof:teo:avg_cap_single_PU}
According to {\it Definition}~\ref{sec:def4}, to evaluate the average of sum capacity of the SU with subcarrier collisions, we have to average {\it a random sum of RVs} with the set of i.i.d. RVs $C_{m,i}^{I,n}$ and $C_{m,i}^{NI}$ as follows:
\begin{equation*}
\begin{split}
\mathbb{E} \left[C_m^1\right] &= \mathbb{E} \left [ \sum_{i=1}^{k_{nm}}C_{m,i}^{I,n}  + \sum_{i=1}^{k_{fm}} C_{m,i}^{NI}   \right ]\\
 &= \mathbb{E} \left [ \mathbb{E} \left [ \sum_{i=1}^{k_{nm}}C_{m,i}^{I,n} \bigg | K_{nm}=k_{nm} \right ] \right ]
\\
& ~~~+ \mathbb{E} \left [ \mathbb{E} \left [ \sum_{i=1}^{k_{fm}}C_{m,i}^{NI} \bigg | K_{fm}=k_{fm} \right ] \right] \\
&= \mathbb{E}\left [  \sum_{i=1}^{k_{nm}}\mathbb{E} \left [ C_{m,i}^{I,n}\right]  \right ]+ \mathbb{E} \left [ \sum_{i=1}^{k_{fm}}\mathbb{E} \left[ C_{m,i}^{NI}\right ]  \right ]
\\
& = \mathbb{E}\left [ k_{nm} \mathbb{E} \left [ C_{m,i}^{I,n}  \right ] \right ]+ \mathbb{E} \left [ k_{fm} \mathbb{E} \left [ C_{m,i}^{NI}  \right ] \right ],
\end{split}
\end{equation*}
where {\it the rule of iterated expectations}~\cite[p.~55,~Theorem 3.24]{wasserman2004all} also known as {\it tower rule}, $\mathbb{E}\left [X\right] = \mathbb{E} \left [ \mathbb{E}\left[X | Y\right]  \right ]$,  is applied, and  the conditional expectations with respect to $k_{nm}\sim \texttt{HYPG}(F_m^S,~F_n^P, ~F) $ and $k_{fm}\sim \texttt{HYPG}(F_m^S,~F- F_n^P, ~F) $   are  used.

Furthermore, $k_{nm}$ and $ C_{m,i}^{I,n}$ are independent, and  so are $k_{fm}$ and $C_{m,i}^{NI}$. Then, we have
\begin{equation*}
\mathbb{E} \left [C_m^1\right] =  \mathbb{E}\left [k_{nm}\right] \mathbb{E}\left  [ C_{m,i}^{I,n}\right]+ \mathbb{E}\left[k_{fm}\right]  \mathbb{E}\left [ C_{m,i}^{NI}\right ].
\end{equation*}

It is also worth to note the relation between the two sums in the first equality that they are independent conditioned with the given $k_{nm}$ and $k_{fm}$ (since $k_{fm} = F_m^S- k_{nm}$).
Taking into account the means of $k_{nm}$ for $n\in [1,N]$ and $k_{fm}$, it follows that $\mathbb{E}\left[k_{nm}\right] = F_m^SF_n^P/F$ and $\mathbb{E}\left[k_{fm}\right] = F_m^S(F- F_n^P)/F$,
which yield the desired result.

\section{Proof of Corollary~\ref{teo:mult_PUs}}\label{proof:teo:mult_PUs}

Following the same approach as in Appendix~\ref{proof:teo:avg_cap_single_PU}, the average capacity in presence of $N$ PUs can be obtained by
using~\eqref{eq:C_m} and the properties of multivariate hypergeometric distribution given in~\eqref{eq:m-hypg} with the means of $k_{nm}$  and $k_{fm} $ expressed as $\mathbb{E}\left[k_{nm}\right] = F_m^SF_n^P/F,~ n=1,\dots,N$ and $\mathbb{E}\left[k_{fm}\right]  = F_m^S\left (F- \sum_{n=1}^{N}F_n^P\right )/F$.

\section{Proof of Corollary~\ref{theo:rate_conv}}\label{proof:theo:rate_conv}
Let $\chi_1=F_n^P \left ( \mathbb{E}\left[C_{m,i}^{I,n}\right] - \mathbb{E}\left[C_{m,i}^{NI}\right] \right )$, $\chi_2= \mathbb{E}\left[C_{m,i}^{NI}\right]$ and
$ C_{m,F}^{avg} =  \mathbb{E} \left[C_m^1\right] = \frac{F_m^S}{F}\chi_1 + F_m^S\chi_2$.
Using {\it Definition}~\ref{sec:def_1}, one can show that
\begin{equation*}
\begin{split}
\lim_{F \to \infty }& \frac{\left | \Delta C_{m,F+1}^{avg}  \right |}{\left | \Delta C_{m,F}^{avg}   \right |}= \\
&  \lim_{F \to \infty } \frac{\left |\frac{F_m^S}{F+2}\chi_1 + F_m^S\chi_2  - \frac{F_m^S}{F+1}\chi_1 - F_m^S\chi_2\right |}{\left | \frac{F_m^S}{F+1}\chi_1 + F_m^S\chi_2  - \frac{F_m^S }{F}\chi_1 - F_m^S\chi_2  \right |}=1,
\end{split}
\end{equation*}
and
\begin{equation*}
\begin{split}
\lim_{F \to \infty } &  \frac{\left |  C_{m,F+1}^{avg} - F_m^S\chi_2\right |}{\left | C_{m,F}^{avg} - F_m^S\chi_2  \right |}  = \\
& \lim_{F \to \infty } \frac{\left |\frac{F_m^S }{F+1}\chi_1 + F_m^S\chi_2  - F_m^S\chi_2\right |}{\left | \frac{F_m^S}{F}\chi_1 + F_m^S\chi_2  -  F_m^S\chi_2  \right |}=1.
\end{split}
\end{equation*}

Hence,  $C_{m,F}^{avg}$ is {\it logarithmically} convergent to $F_m^S \mathbb{E} \left[ C_{m,i}^{NI}\right ]$ as $F \to \infty$.

\section{Evaluation of Limit in Equation~\eqref{eq:limssss}}\label{app:lims}
In the evaluation steps, given in \eqref{eq:lim_eval}, since $f_{C_m^1}(x)\to 0$ and $F_{C_m^1}(x)\to 1$ as $x\to \infty$, first, L'Hopital's rule is applied, and then due to the uniform convergence and the positive terms, the interchange of {\it limit} and {\it infinite sum} is viable. Lastly, because the resulting  expression is of polynomial type, only the highest-order terms are considered.
\begin{figure*}
\begin{equation}\label{eq:lim_eval}
\begin{split}
 \lim_{x \to \infty}\frac{1- F_{C_m^1}(x) }{f_{C_m^1}(x)}
&= \lim_{x \to \infty}\frac{1 - \mathcal{Q} \sum\limits_{k=0}^{\infty}\delta_k  \mathcal{P}\left ( \Delta+k,\frac{x}{\hat{\beta}_{\min}} \right )}
{ \mathcal{Q} \sum\limits_{k=0}^{\infty} \delta_k \frac{ x^{\left ({\Delta+k-1} \right )}  e^{ -x/\hat{\beta}_{\min}}}  {\hat{\beta}_{\min}^{\Delta+k}~\Gamma\left ( \Delta +k \right )}U(x)}
 =  \lim_{x \to \infty}\frac{1 - \mathcal{Q} \sum\limits_{k=0}^{\infty}\delta_k  \left [ 1 - \frac{\Gamma\left ( \Delta +k, \frac{x}{\hat{\beta}_{\min} } \right )}{\Gamma\left ( \Delta +k \right )} \right ] }
{ \mathcal{Q} \sum\limits_{k=0}^{\infty}  \frac{\delta_k x^{{\Delta+k-1} }  e^{ -x/\hat{\beta}_{\min}}}{\hat{\beta}_{\min}^{\Delta+k}~\Gamma\left ( \Delta +k \right )}U(x)} \\
&=  \lim_{x \to \infty} \left [ \lim_{l_k \to \infty}\frac{- \mathcal{Q} \sum\limits_{k=0}^{l_k}\delta_k \frac{x^{{\Delta+k-1}}  e^{ -x/\hat{\beta}_{\min}}}{\hat{\beta}_{\min}^{\Delta+k}~\Gamma\left ( \Delta +k \right )} }
{ \mathcal{Q} \sum\limits_{k=0}^{l_k}\frac{\delta_k ~ e^{-x/\hat{\beta}_{\min}}} {\hat{\beta}_{\min}^{\Delta+k}~\Gamma\left ( \Delta +k \right )}  \left [({\Delta+k-1})x^{{\Delta+k-2} }  - \frac{x^{{\Delta+k-1} }}{\hat{\beta}_{\min}} \right ] }  \right ]
\\
&=  \lim_{x \to \infty} \left [ \lim_{l_k \to \infty}\frac{-  \frac{ \delta_{l_k}x^{{\Delta+l_k-1}}  }{\hat{\beta}_{\min}^{\Delta+l_k}~\Gamma\left ( \Delta +l_k \right )} }
{  \frac{\delta_{l_k} } {\hat{\beta}_{\min}^{\Delta+l_k}~\Gamma\left ( \Delta +l_k \right )}  \left [({\Delta+l_k-1})x^{{\Delta+l_k-2} }  - \frac{x^{{\Delta+l_k-1} }}{\hat{\beta}_{\min}} \right ] } \right ]
\\
&= \lim_{x \to \infty}  \lim_{l_k \to \infty} \frac{-   x^{{\Delta+l_k-1}}  }
{   ({\Delta+l_k-1})x^{{\Delta+l_k-2} }  - \frac{x^{{\Delta+l_k-1} }}{\hat{\beta}_{\min}}  }= \hat{\beta}_{\min}>0.
\end{split}
\end{equation}
\vspace*{0.3cm}
\\
\rule[0.2cm]{1\textwidth}{0.017cm} 
\end{figure*}

\section{Proof of Proposition~\ref{teo:PMF}}\label{sec:PMF}
Recall from \eqref{eq:m-hypg} that the PMF of first SU is given by
\begin{equation*}
p(\mathbf{k}_1)
= \binom{F_f}{k_{f1}}
\binom{F}{F_1^S}^{-1} \prod_{n=1}^{N}\binom{F_n^P}{k_{n1}}.
\end{equation*}

Assuming the orthogonality between subcarriers, given $\mathbf{k}_1$ the conditional PMF of second SU is a multivariate hypergeometric distribution,  is described by
\begin{equation*}
p(\mathbf{k}_2\big |\mathbf{k}_1)
= \binom{F_f - k_{f1}}{k_{f2}}\binom{F -\mathbf{1}^\mathrm{T}\mathbf{k}_1}{F_2^S}^{-1} \prod_{n=1}^{N}\binom{F_n^P - k_{n1}}{k_{n2}},
\end{equation*}
where $\mathbf{1}^\mathrm{T} = [1,1,\dots, 1]^\mathrm{T} \in \mathbb{Z}^{N+1} $ and $\mathbf{1}^\mathrm{T}\mathbf{k}_1 = \sum_{n=1}^{N}k_{n1} + k_{f1}=F_1^S$.
Similarly, for the third SU the conditional PMF for the number of subcarrier collisions is
\begin{equation*}
\begin{split}
p \left ( \mathbf{k}_3\big |\mathbf{k}_1, \mathbf{k}_2 \right )
&= \binom{F_f -  k_{f1}- k_{f2}}{k_{f3}}\binom{F -\mathbf{1}^\mathrm{T}(\mathbf{k}_1 +\mathbf{k}_2)}{F_3^S}^{-1} \\
&~~~\times \prod_{n=1}^{N}\binom{F_n^P - k_{n1} -k_{n2}}{k_{n3}}.
\end{split}
\end{equation*}

In general, for the $m$th SU the conditional PMF is
\begin{equation*}
\begin{split}
p & \left ( \mathbf{k}_m\big |\mathbf{k}_1, \mathbf{k}_2, \dots, \mathbf{k}_{m-1} \right )
= \binom{F_f -  \sum_{j=1}^{m-1}k_{fj}}{k_{fm}}
\\
& \times \binom{F -\mathbf{1}^\mathrm{T}\left(\sum_{j=1}^{m-1}\mathbf{k}_j\right )}{F_m^S}^{-1} \prod_{n=1}^{N}\binom{F_n^P - \sum_{j=1}^{m-1}k_{nj}}{k_{nm}}.
\end{split}
\end{equation*}

Using the chain rule and factorization of PMFs, the joint PMF for SUs is expressed as
\begin{equation*}
p \left (\mathbf{k}_1, \mathbf{k}_2, \dots, \mathbf{k}_{m} \right )
= \prod_{r=2}^{m}p \left (\mathbf{k}_r| \mathbf{k}_{r-1}, \mathbf{k}_{r-2}, \dots, \mathbf{k}_{1} \right ) ~p \left (\mathbf{k}_1  \right ).
\end{equation*}

Finally, the marginal PMF of the $m$th SU with multiple $N$ PUs can be obtained. Based on the evaluations above, it is straightforward to obtain the expected value of $k_{nm}$. Therefore, it is omitted for brevity.


\section*{Acknowledgment}

The authors thank the anonymous reviewers for their valuable comments and suggestions to improve this paper.

\end{document}